\documentclass[11pt,a4paper]{article}
\pdfoutput=1
\usepackage{jcappub}

\usepackage{appendix}
\usepackage{array}
\newcolumntype{x}[1]{>{\centering\arraybackslash}p{#1}}

\usepackage{epsfig}
\usepackage{graphicx}
\usepackage{dcolumn}
\usepackage{amsmath}
\usepackage{enumerate}
\def\lsim{\mathrel{\hbox{\rlap{\hbox{\lower4pt\hbox{$\sim$}}}\hbox{$<$}}}}

\makeatletter

\newcommand{\Rmnum}[1]{\expandafter\@slowromancap\romannumeral #1@}
\makeatother

\title{
Asymmetric dark matter annihilation as a test of non-standard cosmologies
}

\author[a]{Graciela B. Gelmini,}
\author[a]{Ji-Haeng Huh}
\author[a]{and Thomas Rehagen}

\affiliation[a]{Department of Physics and Astronomy, UCLA,\\
475 Portola Plaza, Los Angeles, CA 90095, USA}

\emailAdd{gelmini@physics.ucla.edu}
\emailAdd{jhhuh@physics.ucla.edu}
\emailAdd{trehagen@physics.ucla.edu}


\abstract{We show that the relic abundance of the minority component of asymmetric dark matter  can be very sensitive to the expansion rate of the Universe  and the temperature of transition between a non-standard pre-Big Bang Nucleosynthesis cosmological phase and the standard radiation dominated phase, if chemical decoupling happens before this transition. In particular, because the annihilation cross section of asymmetric dark matter is typically larger than that of symmetric dark matter in the standard cosmology, the decrease in relic density of the minority component in non-standard cosmologies with respect to the majority component may be compensated by the increase in annihilation cross section, so that the annihilation rate at present of asymmetric dark matter, contrary to general  belief, could be larger than that of  symmetric dark matter in the standard cosmology.   Thus, if the annihilation cross section of the asymmetric dark matter candidate is known, the annihilation rate at present, if detectable,  could be used to test the Universe before Big Bang Nucleosynthesis, an epoch from which we do not yet have  any data.}


\begin{document}

\maketitle

\section{Introduction}

The nature of dark matter (DM) is one of the fundamental problems of physics and cosmology.  Particles with weakly interacting cross sections and masses in the few GeV to  10 TeV range, WIMPs (Weakly Interacting Massive Particles), 
are  among the best motivated  DM candidates. DM particle candidates, such as  WIMPs (but also sterile neutrinos and axions)  are produced before Big Bang Nucleosynthesis (BBN),  an epoch from which we have no data.  BBN is the earliest episode (finishing 200 seconds  after the Bang,  when the temperature of the Universe is  T $\simeq 0.8$ MeV) from which we have a trace, the abundance of light elements D, $^4$He and $^7$Li.   
 In order for BBN and all the subsequent  history of the Universe to proceed as usual, it is enough that the earliest and highest temperature during the last radiation dominated period, the so called reheating temperature $T_{RH}$, is larger than 3.2 MeV~\cite{Hannestad:2004px}.  
 
 The argument showing that WIMPs  are good DM candidates, many times called the ``WIMP miracle", is more than 30 years old~\cite{Lee:1977ua}. The density per comoving volume of non-relativistic  particles in thermal equilibrium in the early Universe decreases exponentially with decreasing temperature, due to the Boltzmann factor, until the reactions which change the particle number become ineffective.  
 At this point, when  the annihilation rate  becomes smaller than the Hubble expansion rate, 
  the WIMP number per comoving volume becomes constant. This  moment of chemical decoupling or freeze-out happens later for larger annihilation cross sections $\sigma$, which produces smaller WIMP densities. If  the Universe is radiation dominated during decoupling and there is no subsequent change of entropy in matter plus radiation, the present relic density is

\begin{equation} 
\Omega_\chi^{\rm std} h^2 \simeq 0.1~ \left( \frac{1.8 \times 10^{-9} {\rm ~GeV^{-2}}}{ {\left< \sigma v \right>}} \right),
 \label{OmegaSTD}
\end{equation}

 which for weak order $\sigma \simeq $ G$^2_F m_\chi^2$  gives the right order of magnitude of the DM density (and a temperature  $T_{f.o.} \simeq m_\chi/20$ at freeze-out for a WIMP $\chi$ of mass $m_\chi$).

    The ``WIMP miracle" argument and the standard computation of relic densities rely  on assuming that radiation domination began before the main epoch of production of the relics, that the entropy of matter and radiation has been conserved during and after this epoch, that WIMPs are produced thermally,  i.e. via interactions with the particles in the plasma, and that there is no significant asymmetry between the WIMP particles and antiparticles.  With these assumptions chemical decoupling happens at $T_{f.o.}\simeq m_{\chi}/20$,
  thus all WIMPs with $m_\chi \geq 80$ MeV decouple at temperatures higher than 4 MeV, when  the content and expansion history of the Universe may differ from  the standard assumptions.

     The relic density, see e.g. Ref.~\cite{Gelmini:2006pw} (and also the relic velocity distribution, see e.g. 
     Ref.~\cite{Gelmini:2006vn}) before structure formation of WIMPs (and other DM candidates, e.g. sterile 
     neutrinos~\cite{Gelmini:2004ah}) depends on the characteristics of the Universe (expansion rate, composition, etc.) before BBN.  If these particles are ever found, they would be the first relics from the pre-BBN epoch that could be studied. Thus we will want to extract as much information about the Universe at the moment these particles decoupled as we can.  
     
     Here we present a new potentially detectable effect that non-standard cosmologies may have on asymmetric dark matter.

\section{Asymmetric dark matter}

The idea of asymmetric DM is almost as old as the ``WIMP miracle" argument: if DM particles and antiparticles have an asymmetry similar to the baryonic asymmetry this could explain why the baryonic and DM relic densities are similar, i.e. differing by a factor of a few and not  by many orders of magnitude. 
 In 1985, S. Nussinov~\cite{Nussinov:1985xr} pointed out  that if  the asymmetry of technibaryons and usual baryons could be similar in the early Universe, their present  number relic density would also be similar, which would mean that  the ratio of their relic densities  would be given by  the ratio of the DM and baryon masses. 
It was then largely assumed then that the DM relic density was the critical density, thus $\Omega_{DM}/ \Omega_{B} \simeq 100$ (for $\Omega_{B} \simeq 0.1$) would imply a ratio of the lightest  neutral ``Technibaryon (TB)" and nucleon mass to be $m_{TB} /$GeV $ \simeq 100$ which was phenomenologically acceptable at the time. This argument does not exactly hold now, because the ratio of dark and visible matter relic densities is only about  a factor of 5, but different versions of Technicolour models have been a fertile framework for  100 GeV-TeV mass  asymmetric DM candidates (see e.g. \cite{Chivukula:1989qb}).

The first model unrelated to Technicolour for ``light" asymmetric DM was proposed in 1986. Gelmini, Hall and  Lin \cite{Gelmini:1986zz} produced several models for 5 to 10 GeV mass DM candidates, called ``cosmions" (X),  which could influence the physics of the Sun.  In one of them  the baryon minus ``cosmion"  number $(B-C)$  is assumed to be conserved and a common origin of an asymmetry in both $B$ and $C$ numbers, $\Delta B=\Delta C$ insures that the baryon and DM asymmetries are  identical, so that
$\Omega_X/\Omega_{B} = m_X/$GeV$ \simeq$ 5 to 10  (which was then  considered enough just to account for the ``galactic dark matter").  The same basic idea, i.e. the conservation of a linear combination of baryon and DM particle number and a shared origin of the baryon and dark matter asymmetry was subsequently  realized in many models  for heavier candidates, in the 10's to 100's GeV mass range (see e.g. Ref.~\cite{OTHER}), until 
Kaplan, Luty and Zurek in 2009~\cite{Kaplan:2009ag} applied it again to candidates in the 5 to 15 GeV mass range, in a paper which generated renewed interest in asymmetric DM models (for a recent review see e.g.~\cite{Davoudiasl:2012uw}). Light asymmetric DM particles not unlike ``cosmions" were recently proposed as a means to solve the current  discrepancy in the predicted composition of the Sun between helioseismological data and the revised Standard Solar Model~\cite{Frandsen:2010zz}.

The ideas we develop in the following apply to asymmetric DM WIMPs of any mass. We are not concerned here with the origin of the asymmetry. We assume only that one was generated before the chemical decoupling of the DM particles $\chi$ and antiparticles $\bar{\chi}$, so that
\begin{equation}
Y_{\chi}-Y_{\bar{\chi}}=A,
\label{Y-Y}
\end{equation}
where $Y_{\chi}=n_{\chi}/s$, $Y_{\bar{\chi}}=n_{\bar{\chi}}/s$, $n_{\chi}$ and $n_{\bar{\chi}}$ are the respective relic number densities,  $s$ is the entropy  density, $s=(2\pi^2/45)g_{\star}T^3$, dominated by  the relativistic degrees of freedom $g_{\star}$ and $A$ is a constant that characterizes the asymmetry.  Here we take $A$ positive so  $\chi$ and  ${\bar{\chi}}$ are respectively the majority and minority components of the DM at present.  Moreover, in the following we will assume that $\chi$ and $\bar{\chi}$ account for the whole of the DM, i.e.
\begin{equation}
\Omega_{\chi} +\Omega_{\bar{\chi}}= \Omega_{DM},
\label{Omegas}
\end{equation}
although the arguments can be easily changed if they account for only a fraction of the DM.

\begin{figure} 
\begin{center}
\includegraphics[width=7.5cm]{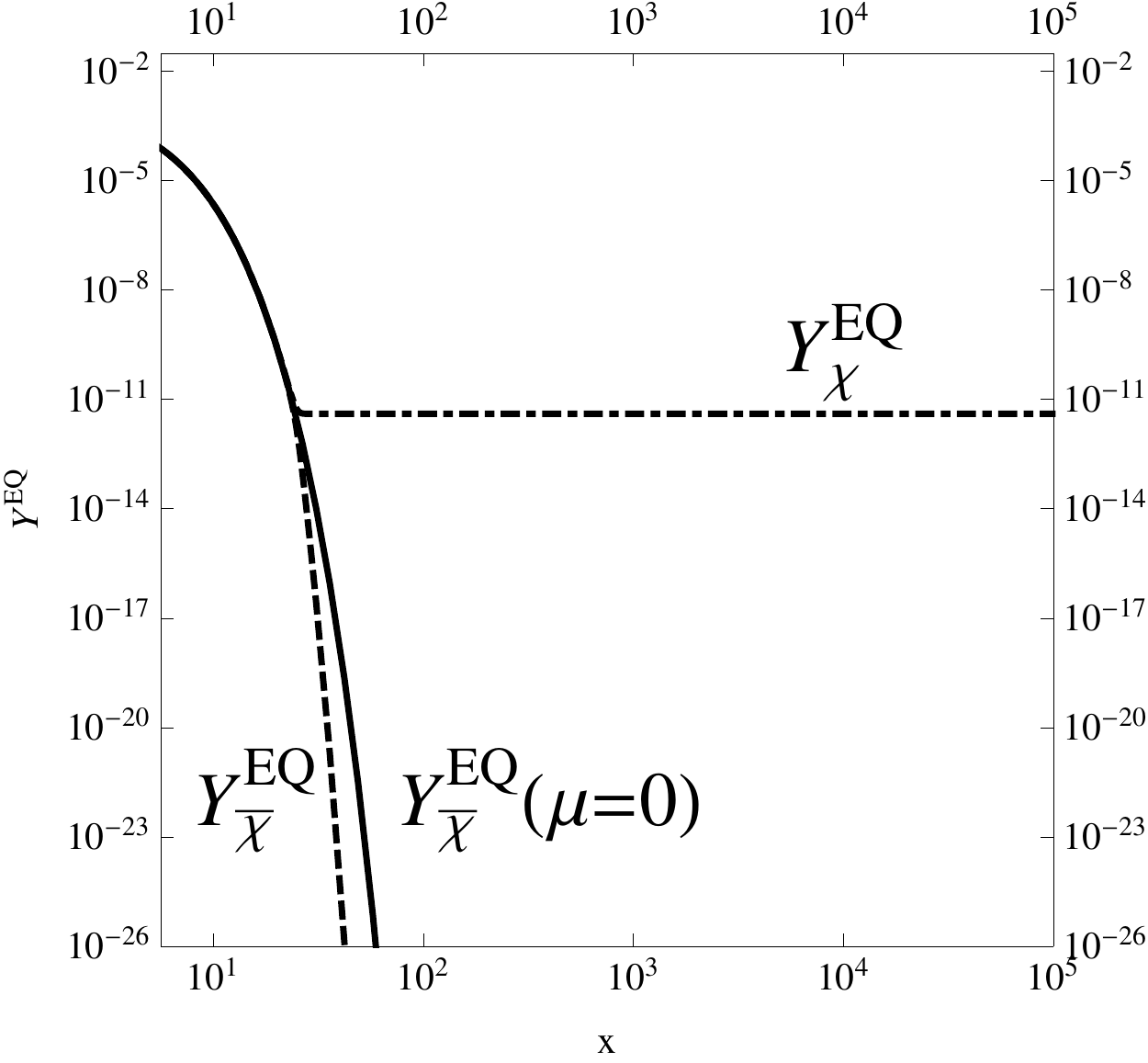}$~~~$
\includegraphics[width=7.5cm]{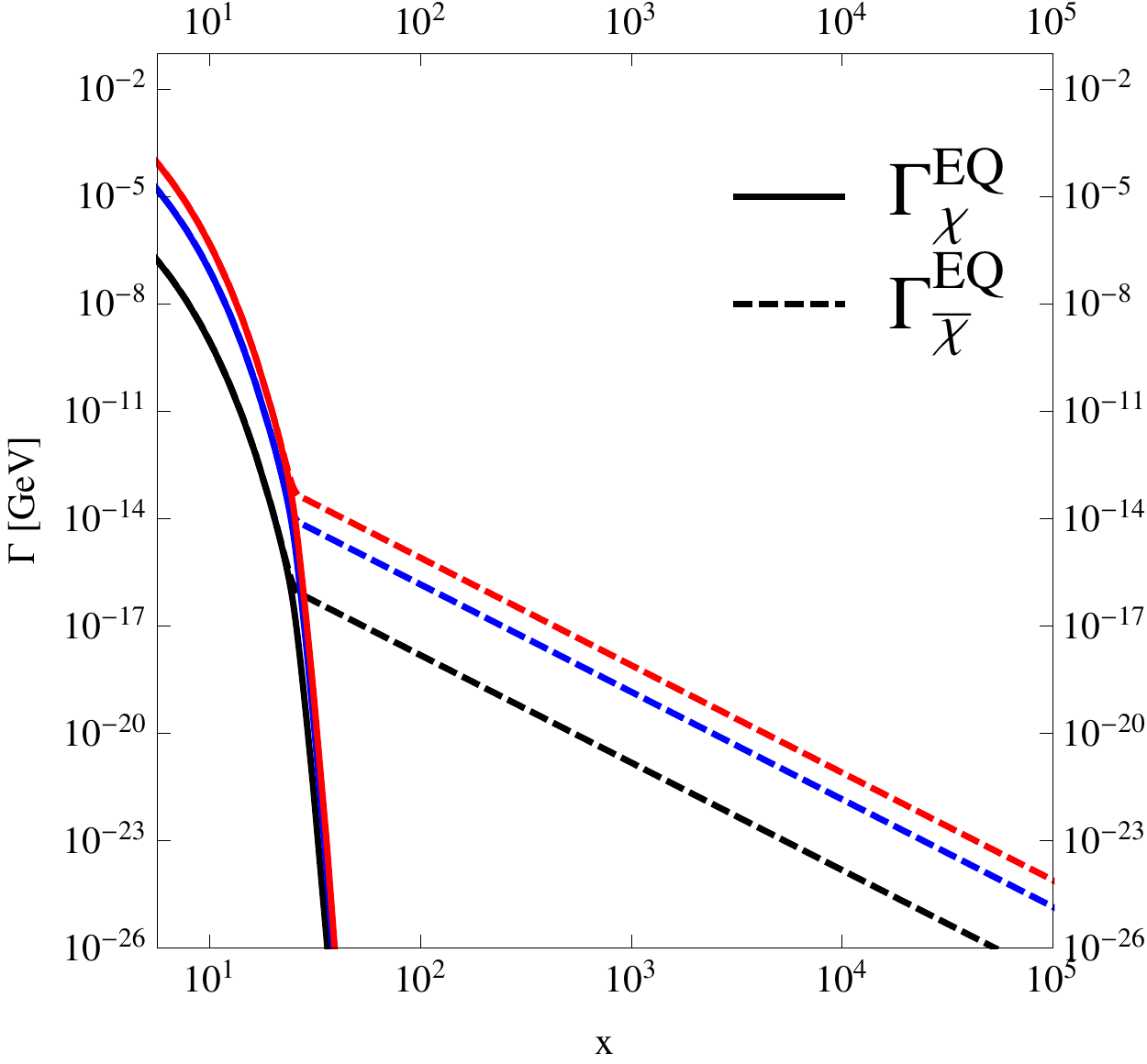}
\caption{1.a (left) Evolution of the equilibrium abundances $Y^{EQ} = n^{EQ}/s$ for the majority, $\chi$, and minority,
 $\bar{\chi}$,  dark matter components, as function of $x= m_\chi/T$ for $A=Y_{\chi}-Y_{\bar{\chi}}=4.05 \times10^{-12}$ and $m_{\chi}=100$ GeV.  Also shown is the equilibrium abundance of symmetric dark matter (A=0).  1.b (right) Equilibrium annihilation rates of $\chi$  and $\bar{\chi}$, $\Gamma_{\chi}^{EQ}$ and $\Gamma_{\bar{\chi}}^{EQ}$  respectively, for three increasing values of the annihilation cross section $ \langle \sigma_{\chi{\bar{\chi}}}  v \rangle$, $9.5\times10^{-9}$GeV$^{-2}$, $9.0\times10^{-7}$GeV$^{-2}$ and $5.0\times10^{-6}$GeV$^{-2}$, for the lower (black), middle (blue) and higher (red) lines respectively.}
\label{new-fig:1}
\end{center}
\end{figure}

The evolution of  the equilibrium values of $Y^{EQ}_{\chi}$, $Y^{EQ}_{\bar{\chi}}$ as function of $x= m_\chi/ T$ are shown in Fig.~\ref{new-fig:1}.a. The equilibrium number densities $n_{\chi}^{EQ}$ and $n_{\bar{\chi}}^{EQ}$ in the presence of an asymmetry differ by the chemical potential  $\mu_{\chi}$ (in equilibrium $\mu_{\chi}=-\mu_{\bar{\chi}}$)
\begin{equation}
n_{\chi}^{EQ}=g_{\chi}\left(\frac{m_{\chi}T}{2\pi}\right)^{3/2}e^{(-m_{\chi}+\mu_{\chi})/T},
\label{nxeq}
\end{equation}
\begin{equation}
n_{\bar{\chi}}^{EQ}=g_{{\chi}}\left(\frac{m_{\chi}T}{2\pi}\right)^{3/2}e^{(-m_{\chi}-\mu_{\chi})/T}.
\label{nxbareq}
\end{equation}
Here, $m_{\chi}$ is the mass of $\chi$ and $\bar{\chi}$, and $g_{\chi}$ is the number of internal degrees of freedom of $\chi$ and of $\bar{\chi}$ separately.  For simplicity in  the following we take $g_\chi = 1$  (which assumes $\chi$ and $\bar{\chi}$ are conjugate complex scalar fields).  A different choice of $g_{\chi}$ (e.g. $g_\chi =2$  if $\chi$ and $\bar{\chi}$ are two conjugate Dirac fermions) would not affect the results significantly.  The results we find would change by factors of $\mathcal{O}(1)$.  The chemical potential can be written in terms of the asymmetry $A$ by substituting  the equilibrium number densities  into Eq.~\eqref{Y-Y}, 
\begin{equation}
n_{\chi}^{EQ}(\mu =0)\left( e^{\mu_\chi /T}-e^{-\mu_\chi /T}\right)=g_{\chi}\left(\frac{m_{\chi}T}{2\pi} \right)^{\frac{3}{2}}e^{- m_{\chi}/T}\left( e^{\mu_\chi /T}-e^{-\mu_\chi /T}\right)=As.
\label{quadratic}
\end{equation}
and solving this quadratic equation for $ e^{\mu_\chi /T}$, to get
\begin{equation}
e^{\mu_\chi /T}=\frac{1}{2}\left(\frac{As}{n_{eq}(\mu =0)}+\sqrt{4+\left(\frac{As}{n_{eq}(\mu =0)}\right)^2}\right).
\label{mut}
\end{equation}

Replacing Eq. \eqref{mut} in Eqs. \eqref{nxeq} and \eqref{nxbareq} we can find $Y_{\chi}^{EQ}=n_{\chi}^{EQ}/s$ and $Y_{\bar{\chi}}^{EQ}=n_{\bar{\chi}}^{EQ}/s$ as functions of $m_\chi$, $A$ and $T$.  See Fig. \ref{new-fig:1}.a for the evolution of $Y_\chi^{EQ}$ and $Y_{\bar{\chi}}^{EQ}$ as a function of $x=m_\chi/T$ for $A=4.05\times 10^{-12}$ and $m_\chi = 100\,{\rm GeV}$.

 In the presence of the asymmetry $A$, the annihilation of the majority component is considerably reduced after $Y_{\chi}$ reaches the value $A$ at $x_A$, $Y_{\chi} (x_A) \simeq A$. For $x>  x_A$ the equilibrium number density per comoving volume of the majority component becomes almost constant    while that of the minority component decreases faster than in the symmetric $A=0$ case as $x$ increases. It is easy to see why this is so by considering the annihilation rate per particle of $\chi$ and $\bar{\chi}$, $\Gamma_{\chi}$ and  $\Gamma_{\bar{\chi}}$  respectively,
\begin{equation}
\Gamma_{\chi} = \langle \sigma_{\chi{\bar{\chi}}} v \rangle n_{\bar{\chi}},~~~~~ \Gamma_{\bar{\chi}} = \langle \sigma_{\chi{\bar{\chi}}} v \rangle n_{\chi}, 
\label{Gammas}
\end{equation}
in equilibrium, whose evolution as function of $x$ is shown in Fig.~\ref{new-fig:1}.b.  Here $\langle \sigma_{\chi{\bar{\chi}}} v \rangle$ is the thermally averaged  $\chi \bar{\chi}$ annihilation cross section and we are assuming that these annihilations are the only processes that can change the number of these particles.

 For $x> x_A$ the $\bar{\chi}$  and ${\chi}$ interaction rate in equilibrium  become respectively larger  and smaller than in the symmetric $A=0$ case because $n_{\chi}$ is much larger and $n_{\bar{\chi}}$ is smaller than in the $A=0$ case.
  
The annihilation of each DM component $\chi$ and $\bar{\chi}$ ceases when their respective interaction rates become smaller than the expansion rate of the Universe $H$, which happens at their respective decoupling or freeze-out, $x_{f o}$ for $\chi$ and $\bar{x}_{f o}$ for $\bar{\chi}$, defined by
\begin{equation}
\Gamma_{\chi} (x_{f o}) = H(x_{f o}),~~~~~ \Gamma_{\bar{\chi}}(\bar{x}_{f o}) = H(\bar{x}_{f o}) .
\label{Freeze-out}
\end{equation}
It is clear from Fig.~\ref{new-fig:1}.b that $\chi$ decouples earlier than $\bar{\chi}$, i.e. $x_{f o}< \bar{x}_{f o}$. At their respective freeze-out, each component acquires its relic density, fixed in comoving volume. However,  little changes in the  abundance of the majority component after it freezes-out, since already at $x_A < x_{f o}$ the annihilations had become very suppressed.

   In the standard cosmology the relic density of the minority component, which was  computed for the first time in Refs.~\cite{Scherrer} and ~\cite{Griest:1986yu} and recently in Refs.~\cite{Iminniyaz:2011cd} and \cite{Graesser}, 
 is exponentially small with respect to the majority component density. (Note that our asymmetry parameter $A$ is called $Q/q$ in Ref.~\cite{Scherrer}, $2\alpha$ in Ref.~\cite{Griest:1986yu}, $C$ in Ref.~\cite{Iminniyaz:2011cd} and $\eta$ in Ref.~\cite{Graesser}.)  This is why there is no asymmetric DM annihilation at present.

Here we will show that if the chemical decoupling of the minority component  happens during a non-standard cosmological phase, the relic abundance of this component of asymmetric dark matter  can be much larger than in the standard cosmology. Thus, if the annihilation cross section of the dark matter candidate is known,
the annihilation rate at present, if detectable,  could be used to test the Universe before Big Bang Nucleosynthesis, an epoch from which we do not  yet have  any data. In particular, because the annihilation cross section of asymmetric dark matter is typically larger than that of symmetric dark matter in the standard cosmology, the decrease in relic density of the minority component with respect to the majority component may be small enough that the annihilation rate at present of asymmetric dark matter, contrary to general  belief, could be larger than that of symmetric dark matter in the standard cosmology.

\section{Non-standard pre-BBN cosmologies}

Many viable non-standard pre-BBN cosmological models have been proposed (see e.g. \cite{Gelmini:2010zh} for a review). Usually non-standard cosmological  scenarios contain additional parameters that can be adjusted to modify the WIMP relic density. However these are due to physics that does not manifest itself in accelerator or DM detection experiments.

  In the standard (STD) cosmology  the Universe is radiation dominated, thus the expansion rate of the Universe $H$ depends on the temperature of the radiation bath 
  $T$ as   $H \sim T^2$, but  in   non-standard models  $H$ can decrease slower, $H \sim T^{1.2}$,  or faster,  $H \sim T^3$
  as $T$ decreases.  This is what happens  in ``scalar-tensor"~\cite{Santiago:1998ae,  Catena:2004ba} and ``kination"~\cite{Salati:2002md} cosmological models, respectively. In both of these models the entropy of matter plus radiation is conserved (and it is dominated by the radiation component), thus the usual relation between $T$ and the scale factor of the Universe $a$, namely $a \sim T^{-1}$, is the same as  for a radiation dominated Universe.
  \begin{figure}
\begin{center}
\includegraphics[width=7.5cm]{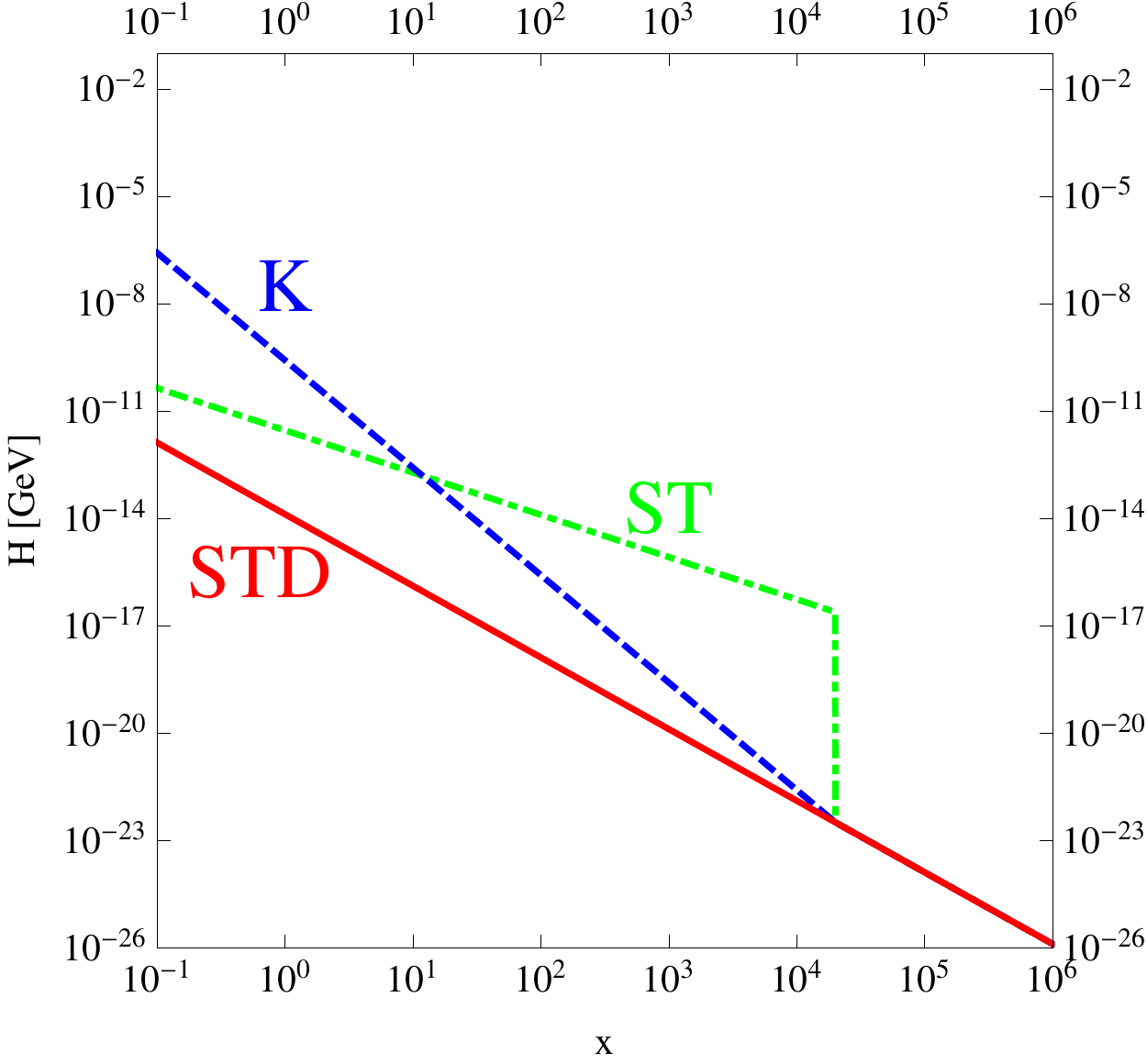}
\caption{ Expansion rate of the Universe $H$ for the standard cosmology, STD (solid red line) and for the kination, K (dashed blue line) and scalar-tensor, ST (dot-dashed green line) cosmologies with a transition temperature $T_{tr}= 5$ MeV to the standard cosmology.}
\label{fig:1}
\end{center}
\end{figure}
 In models in which the potential energy of a scalar field  oscillating  around its true minimum while decaying is the dominant component of the Universe just before BBN,  the expansion rate decreases even faster, $H \sim T^4$ but there is entropy creation, thus $T \sim a^{-3/8}$. These models produce the largest departure of symmetric WIMP relic density from the standard density (see e.g. \cite{Gelmini:2006pw})  and we could expect their effect on asymmetric DM to be very large too. However here for simplicity we concentrate on models  in which there is no entropy creation.
  
Scalar-tensor (ST) theories of gravity~\cite{Santiago:1998ae, Catena:2004ba} incorporate a scalar field coupled only through the metric tensor to the matter fields. In many of these models the expansion of the Universe  drives the scalar field towards a state where the theory is indistinguishable from General Relativity at a transition temperature $T_{tr}$, but  the effect of the scalar field changes the expansion rate of the Universe at earlier times, either increasing or decreasing it. Theories with a single matter sector  typically predict an enhancement of $H$ before BBN. In Ref.~\cite{Catena:2004ba}, for large temperatures $T> T_{tr}$ the $H$ is enhanced  with respect to the standard expansion rate $H_{\rm STD}$ by a factor  $f_\phi \simeq 2.19 \times 10^{14} (T_0/T)^{0.82}$ ($T_0$ is the present temperature of the Universe),  thus   $H_{ST} \sim T^{1.2}$. At $T_{tr}$,  $f_\phi$ drops sharply to values close to 1 at $T < T_{tr}$ before BBN sets is.

  Kination~\cite{Salati:2002md} is a period  in which the kinetic energy $\rho_\phi\simeq  \dot{\phi}^2/2$ of a scalar field $\phi$ (maybe a quintessence field) dominates over the potential energy so $\rho_{\rm total} \simeq  \dot{\phi}^2/2 \sim a^{-6}$. Since $T\sim a^{-1}$ as usual,  $H \sim \sqrt{\rho_{\rm total} } \sim T^3$. The contribution of the $\phi$ kinetic energy to the total density is usually quantified through the ratio of $\phi$ -to-photon energy density, $\eta_\phi = \rho_\phi/ \rho_\gamma$ at $T\simeq 1$ MeV so that at higher temperatures $H\simeq \sqrt{\eta_\phi} (T/ 1{\rm  MeV}) H_{\rm STD}$. Notice that at $T\simeq 1$ MeV, i.e. during BBN, the quintessence field cannot be dominant, thus $\eta_\phi <1$.
  
Fig.~\ref{fig:1} shows the standard  (STD) expansion rate, and the scalar-tensor (ST) and kination (K) expansion rates for a transition temperature ($T_{tr}$ is the temperature at which the cosmology becomes standard)  $T_{tr}= 5$ MeV.  As mentioned in the introduction, the lower bound on $T_{tr}$ is $3.2 \, {\rm MeV}$ \cite{Hannestad:2004px}.

\section{Asymmetric DM relic density calculation}

The relic densities of DM particles $\chi$ and anti-particle $\bar{\chi}$ are calculated by solving their respective Boltzmann equation (for asymmetric DM, $\chi$ and  $\bar{\chi}$ are not self-conjugate,  $\chi \neq \bar{\chi}$). We assume that close to the moment of decoupling, the only reactions that  change the number of DM particles and antiparticles are annihilations of   $\chi\bar{\chi}$ pairs into Standard Model particles and the inverse process of pair creation. With this assumption we have
\begin{equation}
\frac{dn_{\chi}}{dt}+3Hn_{\chi}=\frac{dn_{\bar{\chi}}}{dt}+3Hn_{\bar{\chi}}=-\langle\sigma_{\chi{\bar{\chi}}} v\rangle(n_{\chi}n_{\bar{\chi}}-n^{EQ}_{\chi}n^{EQ}_{\bar{\chi}}).
\label{dndt}
\end{equation}
Notice that in \eqref{dndt} we have disregarded the possibility of having $\chi \chi$ and $\bar{\chi} \bar{\chi}$  self-annihilation, which is instead considered in Ref. \cite{Ellwanger}.
In terms of the dimensionless quantities $Y_{\chi}$, $Y_{\bar{\chi}}$ and $x$, and treating $g_{\star}$ as a constant, the Boltzmann equations become
\begin{equation}
\frac{dY_{\chi}}{dx}=\frac{dY_{\bar{\chi}}}{dx}=\frac{-\langle\sigma_{\chi{\bar{\chi}}} v\rangle}{H}\frac{2\pi^2}{45}g_{\star}\frac{m_{\chi}^3}{x^4}(Y_{\chi}Y_{\bar{\chi}}-Y^{EQ}_{\chi}Y^{EQ}_{\bar{\chi}}),
\label{dYdx}
\end{equation}
where (see Eqs.~\eqref{nxeq} and \eqref{nxbareq})
\begin{equation}
Y_{\chi}^{EQ}Y_{\bar{\chi}}^{EQ}=\frac{1}{(2\pi)^3}\left(\frac{45}{2\pi^2}\right)^2\left(\frac{g_{\chi}}{g_{\star}}\right)^2x^3e^{-2x}.
\label{YYEQ}
\end{equation}
In the following we take  $g_{\star}=90$, a good approximate value for decoupling temperatures above the QCD phase transition.  Eq.~\eqref{dYdx} implies $d(Y_{\chi}-Y_{\bar{\chi}})/ dx=0$, so the asymmetry $A$  defined in Eq.~\eqref{Y-Y} is a constant.  It is clear that $A$ is constant because we are only considering interactions of the form  $\chi\bar{\chi}\leftrightarrow p$'s  where $p$'s are Standard Model particles, which leave the difference between the co-moving number density of the particle and anti-particle unchanged.  We are assuming that the asymmetry has been produced prior to the epoch that we are considering.

Using Eq.~\eqref{Y-Y}, we finally obtain the equations we wish to solve
\begin{equation}
\frac{dY_{\chi}}{dx}=\frac{-\langle\sigma_{\chi{\bar{\chi}}}  v\rangle}{H}\frac{2\pi^2}{45}g_{\star}\frac{m_{\chi}^3}{x^4}(Y_{\chi}^2-AY_{\chi}-Y^{EQ}_{\chi}Y^{EQ}_{\bar{\chi}}),
\label{BY}
\end{equation}
\begin{equation}
\frac{dY_{\bar{\chi}}}{dx}=\frac{-\langle\sigma_{\chi{\bar{\chi}}}  v\rangle}{H}\frac{2\pi^2}{45}g_{\star}\frac{m_{\chi}^3}{x^4}(Y_{\bar{\chi}}^2+AY_{\bar{\chi}}-Y^{EQ}_{\chi}Y^{EQ}_{\bar{\chi}}).
\label{BYbar}
\end{equation}

\subsection{Analytical solutions}

It is useful to obtain approximate analytical solutions to Eq.~\eqref{BYbar} and compare them with  numerical solutions. Fig.~\ref{new-fig:1}.b makes it clear that in the period between $x_A$, when $Y_\chi$ becomes practically constant and the freeze-out $\bar{x}_{f o}$ of the minority component, $Y_{\bar{\chi}}$ decreases exponentially. Until  $\bar{\chi}$ freezes out, the densities of both DM components are those of equilibrium (for $\chi$ it is very close) and after, for $x>\bar{x}_{fo}$,  the production term in the Boltzmann equations is suppressed compared to the annihilation term (because $Y_{\bar{\chi}} \ll Y^{EQ}_{\bar{\chi}}$) and so we ignore it.  Then, for $x>\bar{x}_{fo}$, Eq.~\eqref{BYbar} becomes
\begin{equation}
\frac{dY_{\bar{\chi}}}{dx}=\frac{-\langle\sigma_{\chi{\bar{\chi}}}  v\rangle}{H}\frac{2\pi^2}{45}g_{\star}\frac{m_{\chi}^3}{x^4}\left(Y_{\bar{\chi}}^2+AY_{\bar{\chi}}\right).
\label{lateBYbar}
\end{equation}
Integrating this equation from $\bar{x}_{fo}$ to $\infty$ (a good approximation for the value of $x$ at present) we get
\begin{equation}
\log\left (\frac{Y_{\bar{\chi}}}{Y_{\bar{\chi}}+A}\middle )\right |_{\bar{x}_{fo}}^{\infty}=-A\int ^\infty _{\bar{x}_{fo}} \frac{\langle\sigma_{\chi{\bar{\chi}}}  v\rangle}{H}\frac{2\pi^2}{45}g_{\star}\frac{m_{\chi}^3}{x^4}dx.
\label{Ybarintl}
\end{equation}

For asymmetric DM the ratio  $Y_{\chi}/ Y_{\bar{\chi}}$ increases  immediately after ${\bar{\chi}}$ decouples, as $Y_{\bar{\chi}}$ decreases.  The decoupling of $\bar{\chi}$ is not instantaneous, as assumed when defining $\bar{x}_{fo}$, and annihilations still continue for a while after $Y_{\bar{\chi}}$ departs from its equilibrium value.  This can be seen e.g. in 
Figs.~\ref{fig:Compare-STD}  and \ref{fig:Compare} where the solid black lines showing the numerical solutions for $Y_{\bar{\chi}}$ decrease after the freeze-out of the minority component at $\bar{x}_{fo}\gtrsim x_A$. So we keep only the $x \rightarrow \infty$ term on the left hand side of Eq.~\eqref{Ybarintl}  and obtain
\begin{equation}
Y_{\bar{\chi}}(x \rightarrow \infty)= A \left[\exp{\left(A\int ^\infty _{\bar{x}_{fo}} \frac{\langle\sigma_{\chi{\bar{\chi}}}  v\rangle}{H}\frac{2\pi^2}{45}
g_{\star}\frac{m_{\chi}^3}{x^4}dx\right)}-1\right]^{-1}.
\label{Ybarsol}
\end{equation}
 Using Eq.~\eqref{Y-Y}, i.e.  $Y_{\chi}(x \rightarrow \infty) = A+ Y_{\bar{\chi}}(x \rightarrow \infty )$ we get for the majority component
\begin{equation}
Y_{\chi}(x \rightarrow \infty)=A \left[1-\exp{\left(-A\int ^\infty _{\bar{x}_{fo}} \frac{\langle\sigma_{\chi{\bar{\chi}}}  v\rangle}{H}\frac{2\pi^2}{45}g_{\star}\frac{m_{\chi}^3}{x^4}dx\right)}\right]^{-1}.
\label{Ysol}
\end{equation}

We have kept $H$ explicitly because we want to explore different possibilities for the Hubble expansion.  Had we replaced $H$ by its functional form in the standard cosmology, these equations would be identical to those in Ref. \cite{Iminniyaz:2011cd}, and the ratio of $Y_{\bar{\chi}}/Y_{\chi}$ resulting from these equations is the same as in Ref. \cite{Graesser}.

\subsubsection{Standard cosmology}

 In the standard cosmology, the expansion rate of the Universe is
\begin{equation}
H_{STD}=\frac{\pi T^2}{M_{P}}\sqrt{\frac{g_{\star}}{90}}
\label{Hstd}
\end{equation}
where $M_{P}$ is the reduced Planck mass, and $g_{\star}$ is the effective number of relativistic degrees of freedom.  Expanding the annihilation cross section in powers of the relative velocity, v, and then taking the thermal average, the annihilation cross section can be written in term of constants $a$ and $b$ as
\begin{equation}
\langle \sigma_{\chi{\bar{\chi}}}  v \rangle=a+3bx^{-1}+\mathcal{O}(x^{-2}).
\label{sigmav}
\end{equation}
 where the $a$ term is dominant for s-wave $\chi \bar{\chi}$ annihilation, and $a=0$ for p-wave annihilation.

Using Eqs.~\eqref{Hstd} and \eqref{sigmav}, we solve the integral in Eq.~\eqref{Ybarsol} with the result
\begin{equation}
Y^{STD}_{\bar{\chi}}(x\rightarrow \infty)=A \left\{\exp\left[A\lambda \left(\frac{a}{\bar{x}_{fo}}+\frac{3b}{2\bar{x}_{fo}^2}\right)\right]-1\right\}^{-1},
\label{Ybarsolstd}
\end{equation}
where  $\lambda$ is defined as $\lambda = -\left[(ds/dx)/3H\right]_{x=1}= 4 \pi m_{\chi}M_{P}\sqrt{g_{\star}/90}=3.0\times 10^{21} m_{100}\, {\rm GeV}^2$, and $m_{100}=m_\chi /$100 GeV. From Eq. \eqref{Ysol}, we get
\begin{equation}
Y^{STD}_{\chi}(x\rightarrow \infty)=A \left\{1-\exp{\left[-A\lambda \left(\frac{a}{\bar{x}_{fo}}+\frac{3b}{2\bar{x}_{fo}^2}\right)\right]}\right\}^{-1}.
\label{Ysolstd}
\end{equation}

\subsubsection{Kination}

At the transition temperature $T_{tr}$ at which the Universe becomes radiation dominated after the kination phase, the expansion rate of the Universe during kination, $H_K$, coincides with the standard one, $H_K(T_{tr}) \simeq H_{STD}(T_{tr})$ which allows us to fix the constant multiplying $T^3$ in $H_K$ and find
\begin{equation}
H_{K}(T) \simeq \frac{\pi}{M_{P}T_{tr}}\sqrt{\frac{g_{\star}}{90}}T^3.
\label{HK}
\end{equation}
This corresponds to having the kination parameter $\sqrt{\eta _{\phi}}=1$MeV$/T_{tr}$, so for $T_{tr} \geq$ 1 MeV,  $\eta _{\phi} \leq 1$ during BBN. 

Using Eq.~\eqref{HK} in Eq.~\eqref{Ybarsol} we calculate the relic abundance of $\bar{\chi}$ due to kination,
\begin{equation}
Y^{K}_{\bar{\chi}}(x\rightarrow \infty)={A}\left\{{\left(\frac{x_{tr}}{\bar{x}_{fo}}\right)^{(A\lambda a/x_{tr})}~\exp\left[\frac{A\lambda a}{x_{tr}}\right]~\exp\left[\frac{3A\lambda b}{x_{tr}}\left(\frac{1}{\bar{x}_{fo}}-\frac{1}{2x_{tr}}\right)\right]-1}\right\}^{-1}
\label{YbarsolK}
\end{equation}
where $x_{tr}=m_{\chi}/T_{tr}$, and $\lambda $ is defined as before.  An expression for the relic abundance of $\chi$ can be found by  using  Eq.~\eqref{Y-Y}, i.e.  $Y_{\chi}(x \rightarrow \infty) = A + Y_{\bar{\chi}}(x \rightarrow \infty )$.

\subsubsection{Scalar-tensor model}

 As mentioned above, in this model~\cite{Catena:2004ba} the expansion rate of the Universe is given by
\begin{equation}
H_{ST}(T)= f_\phi(T)H_{STD}(T)
\label{HST}
\end{equation}
where $f_\phi(T)\simeq 2.19\times 10^{14}\left({T_0}/{T}\right)^{0.82}$ is a temperature dependent factor which transitions very fast to a value of 1 at the transition temperature $T_{tr}$, and $T_0$ is the present temperature of the Universe.  In terms of $x$, this factor is $f_\phi(x)\simeq 9.65\times 10^3\left({\rm GeV}/m\right)^{0.82}x^{0.82}$. Thus,
for transition temperatures from 1 MeV to a few GeV,  the Hubble parameter decreases suddenly by a few orders of magnitude at $T_{tr}$, when the change to the standard cosmology happens.  Due to this sudden drop in the $H$, $\chi$ and $\bar{\chi}$ start re-annihilating and they freeze-out again shortly after.  Usually $\chi$ and $\bar{\chi}$ freeze-out  in the re-annihilation phase before they can reach equilibrium and, in this case, we can still neglect the production term in the Boltzmann equations (again because $Y_{\bar{\chi}}\gg Y^{EQ}_{\bar{\chi}}$), and Eqs.~\eqref{Ybarsol} and \eqref{Ysol} still hold.  Using Eq.~\eqref{HST} in Eq.~\eqref{Ybarsol}, we calculate the relic abundance of $\bar{\chi}$ in the scalar-tensor cosmology,
\begin{equation}
Y^{ST}_{\bar{\chi}}(x\rightarrow \infty)=A\left\{{\rm exp}\left[A\lambda a \left(\frac{1}{x_{tr}}+5.69 \times 10^{-5}\left( \frac{m_\chi}{1 {\rm GeV}}\right) ^{0.82}\left(\bar{x}_{fo}^{-1.82}-x_{tr}^{-1.82}\right)\right)\right]-1\right\}^{-1}
\label{YbarsolST}
\end{equation}
where $x_{tr}=m_{\chi}/T_{tr}$, and $\lambda $ is defined as before.  We have taken $b=0$ for simplicity.  As  before, an expression for the relic abundance of $\chi$ can be found by  using  Eq.~\eqref{Y-Y}, i.e.  $Y_{\chi}(x \rightarrow \infty) = A + Y_{\bar{\chi}}(x \rightarrow \infty )$.

\subsubsection{Estimation of the $\bar{\chi}$ freeze-out temperature}

To find $\bar{x}_{fo}$ we use the freeze-out condition Eq.~\eqref{Freeze-out} for $\bar{\chi}$, using in each case the corresponding $H$ written above and the reaction rate of $\bar{\chi}$, $\Gamma_{\bar{\chi}}$.  Considering that until freeze-out  $n_{\chi}$ and $n_{\bar{\chi}}$  closely track their  equilibrium values, $n_{\chi}^{EQ}$ and $n_{\bar{\chi}}^{EQ}$ given  in Eqs.~\eqref{nxeq}  and \eqref{nxbareq}, we can approximate $\Gamma_{\bar{\chi}}$ by
\begin{equation}
\Gamma_{\bar{\chi}}\simeq \langle \sigma_{\chi{\bar{\chi}}}  v \rangle n_{\chi}^{EQ}(\mu=0)e^{\mu_{\chi}/T}.
\label{Gammachi}
\end{equation}
Using Eqs.~\eqref{quadratic} and \eqref{mut} we get an expression for $\Gamma_{\bar{\chi}}$ as function of the asymmetry $A$. Using this form of $\Gamma_{\bar{\chi}}$,  the $\bar{\chi}$ freeze-out condition Eq.~\eqref{Freeze-out} becomes a transcendental equation for  $\bar{x}_{f o}$, which we can solve numerically. After solving for $\bar{x}_{f o}$,
  we use the expressions calculated above to estimate the relic abundance of $\chi$ and $\bar{\chi}$.

\subsection{Numerical solutions}

The Boltzmann equations ~\eqref{BY}  and \eqref{BYbar} can be solved numerically for each of the cosmologies we consider.  Here we show examples of the evolution of $Y_{\chi}$ and $Y_{\bar{\chi}}$ as a function of $x$ in each cosmology and later compare the numerically calculated relic density to the analytical results.  In all of the following we present results for pure s-wave $\chi \bar{\chi}$ annihilation.

The final relic density of the minority component depends strongly on the relation between its  freeze-out temperature or $\bar{x}_{f o}$ and the value  $x_A$ at which  the asymmetry becomes important and the minority and majority densities  $Y_{\chi}^{EQ}$ and $Y_{\bar{\chi}}^{EQ}$ become very different, given by  $Y_\chi^{EQ}(x_A) \simeq A$ (see 
in Fig.~\ref{fig:1}.a).  

One can distinguish three cases. The first is when ${\bar{x}}_{f o}  < x_A$, i.e. the decoupling of $\bar{\chi}$, and also $\chi$ since in this case ${x}_{f o} = {\bar{x}}_{f o}$, happens  before the split of  $Y_{\chi}^{EQ}$ and $Y_{\bar{\chi}}^{EQ}$, then  $Y_{\chi}\simeq Y_{\bar{\chi}}$ at freeze-out, and so the present relic abundance of $\chi$ and $\bar{\chi}$ are practically the same.  This case is very similar to that of symmetric DM.

The second case is when  $\bar{x}_{f o} \gg x_A$, i.e. when the decoupling of $\bar{\chi}$ occurs  well after the split between  $Y_{\chi}^{EQ}$ and  $Y_{\bar{\chi}}^{EQ}$, so that  $Y_{\bar{\chi}}^{EQ}(\bar{x}_{f o})$ is exponentially smaller than   $Y_{\chi} \simeq A$, thus the $\bar{\chi}$ present  relic density is completely negligible when compared to the $\chi$ density.

\begin{figure}
\begin{center}$
\begin{array}{cc}
\includegraphics[width=7.5cm]{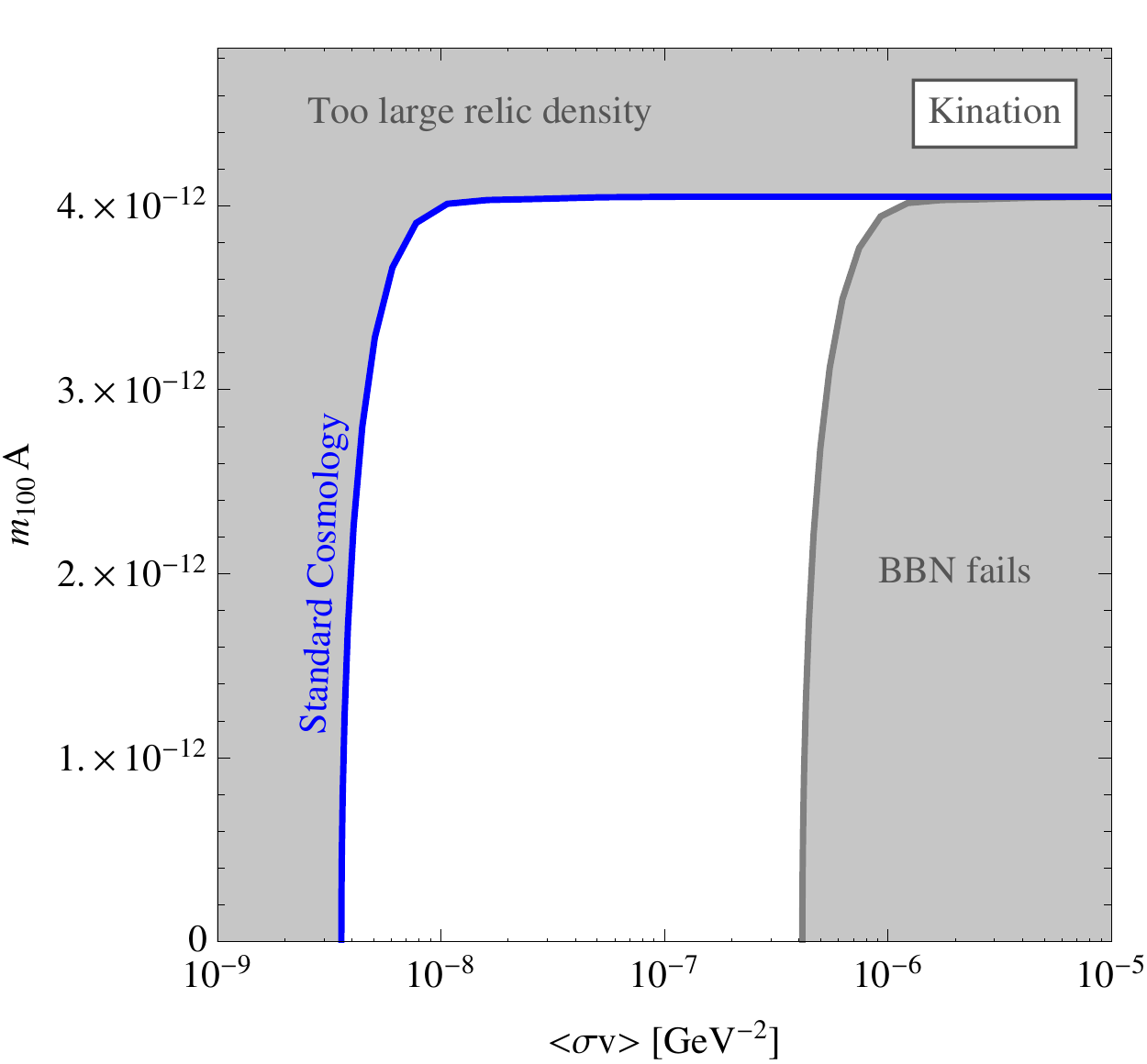}&
\includegraphics[width=7.5cm]{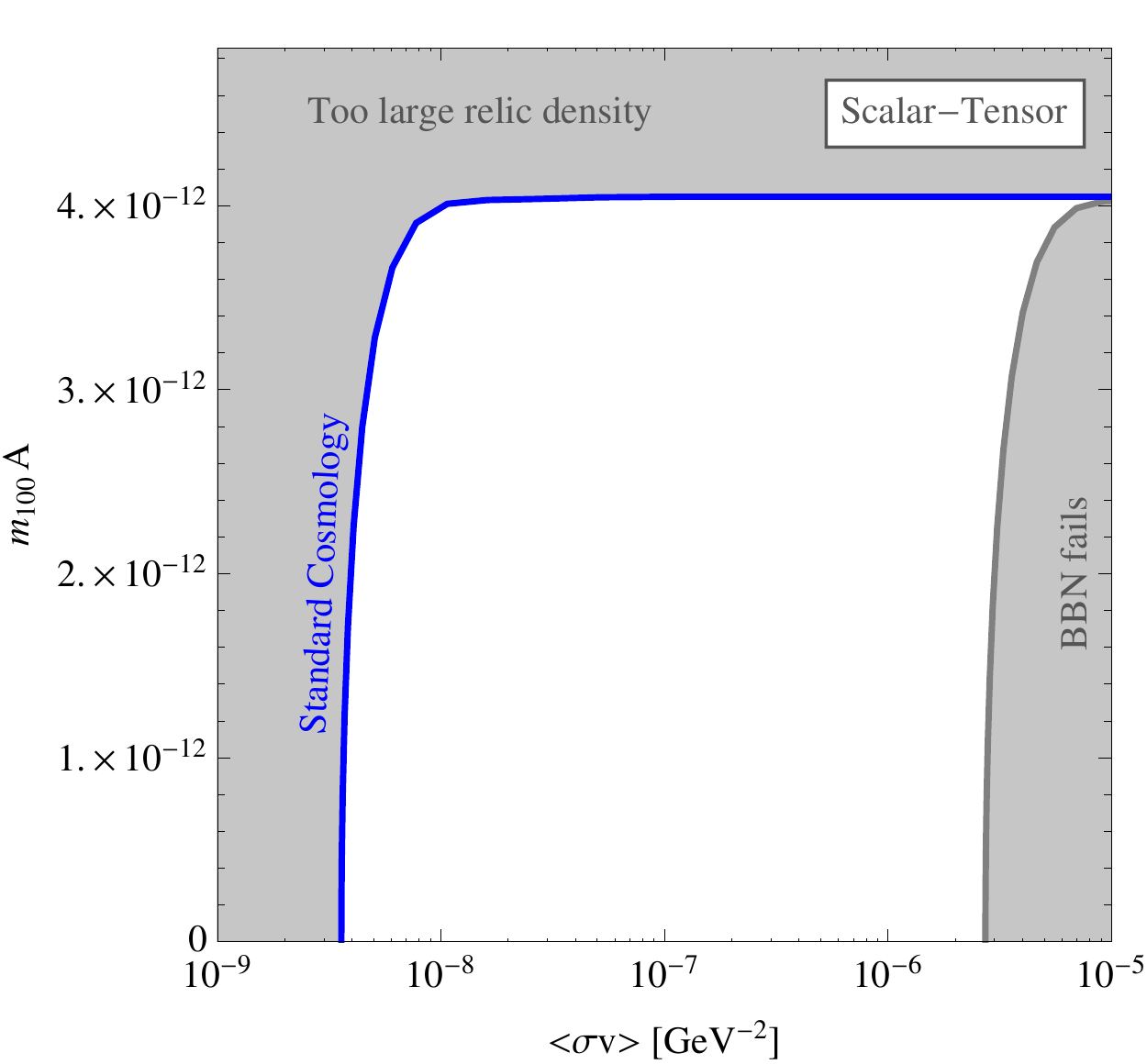}
\end{array}$
\caption{Values of $m_{100}A$ for which $\chi$ and ${\bar{\chi}}$ constitute all of the DM, as function of the $\chi {\bar{\chi}}$ annihilation cross section $\langle \sigma v \rangle = \langle \sigma_{\chi \bar{\chi}} v \rangle$, in the standard cosmology (blue contour in both panels) and \ref{new-fig:6}.a (left) kination and  \ref{new-fig:6}.b (right) scalar-tensor models. The solid gray contour to the right in each plot corresponds to a transition temperature of $T_{tr}=5$ MeV, and the blue contour has a transition temperature equal to the standard cosmology freeze-out temperature, $T_{fo}\simeq m_{100}5\, {\rm GeV}$.  The contours depends very slightly on $m_\chi$  for the standard cosmology, and  if  $m_\chi > 100$ GeV also  for the other two (for $m_\chi \simeq 10$ GeV and   $T_{tr}$ close to 5 MeV, the contours change appreciably with respect to those shown here- see Fig.~\ref{fig:Mass}).  }
\label{new-fig:6}
\end{center}
\end{figure}

\begin{figure}
\begin{center}
\includegraphics[width=7.5cm]{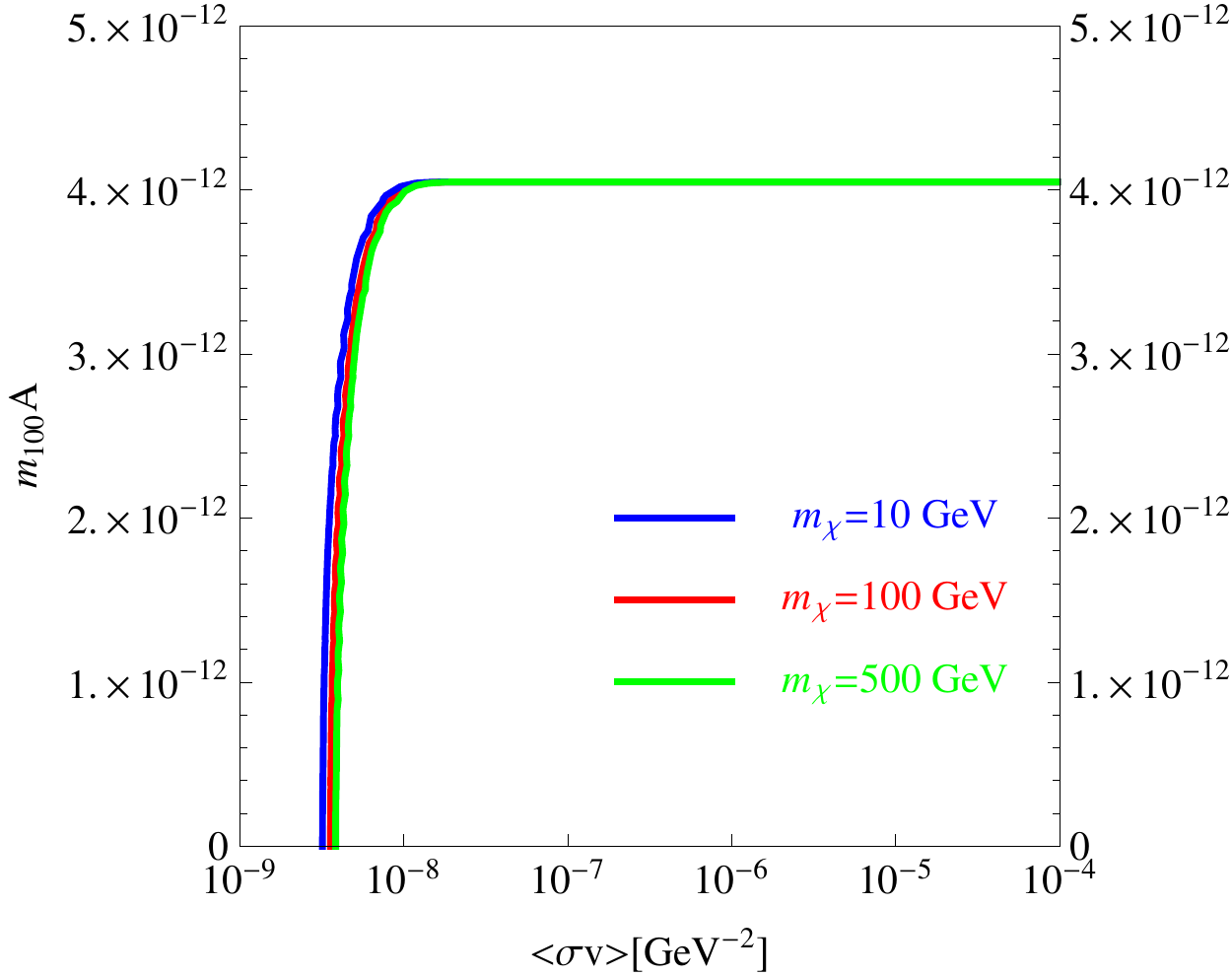}
\caption{Contours of  $m_{100}A$ for which $\chi$ and $\bar{\chi}$ make up all of the DM as a function of the $\chi{\bar{\chi}}$ annihilation cross section $\langle \sigma v \rangle =\langle \sigma_{\chi \bar{\chi}} v \rangle$ for $m_\chi = 10\, {\rm GeV}$ (green), $m_\chi = 100\, {\rm GeV}$ (red), and $m_\chi = 500\, {\rm GeV}$ (blue) in the standard cosmology.}
\label{fig:Mass-STD}
\end{center}
\end{figure}

\begin{figure}
\begin{center}
\includegraphics[width=7.5cm]{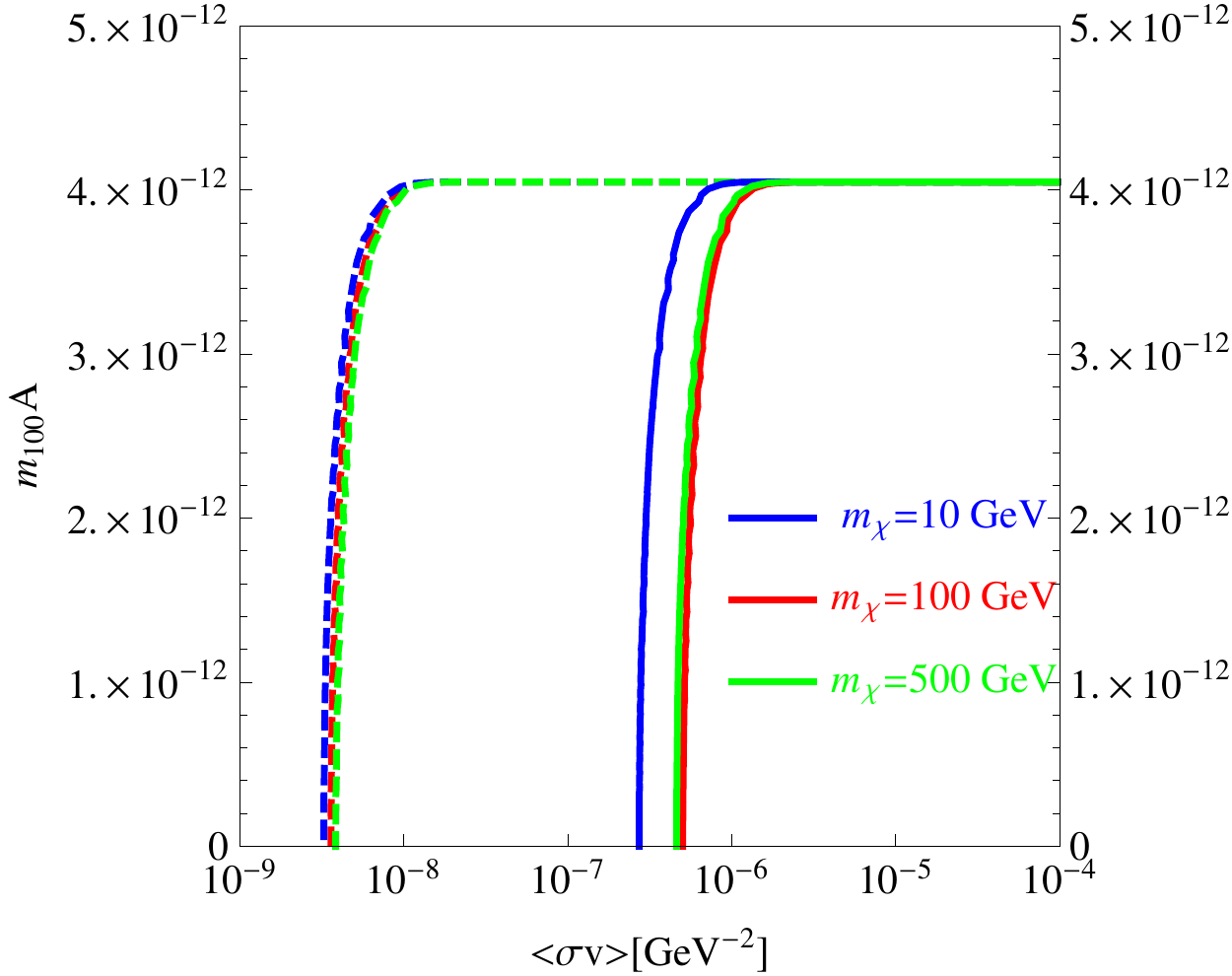}
\includegraphics[width=7.5cm]{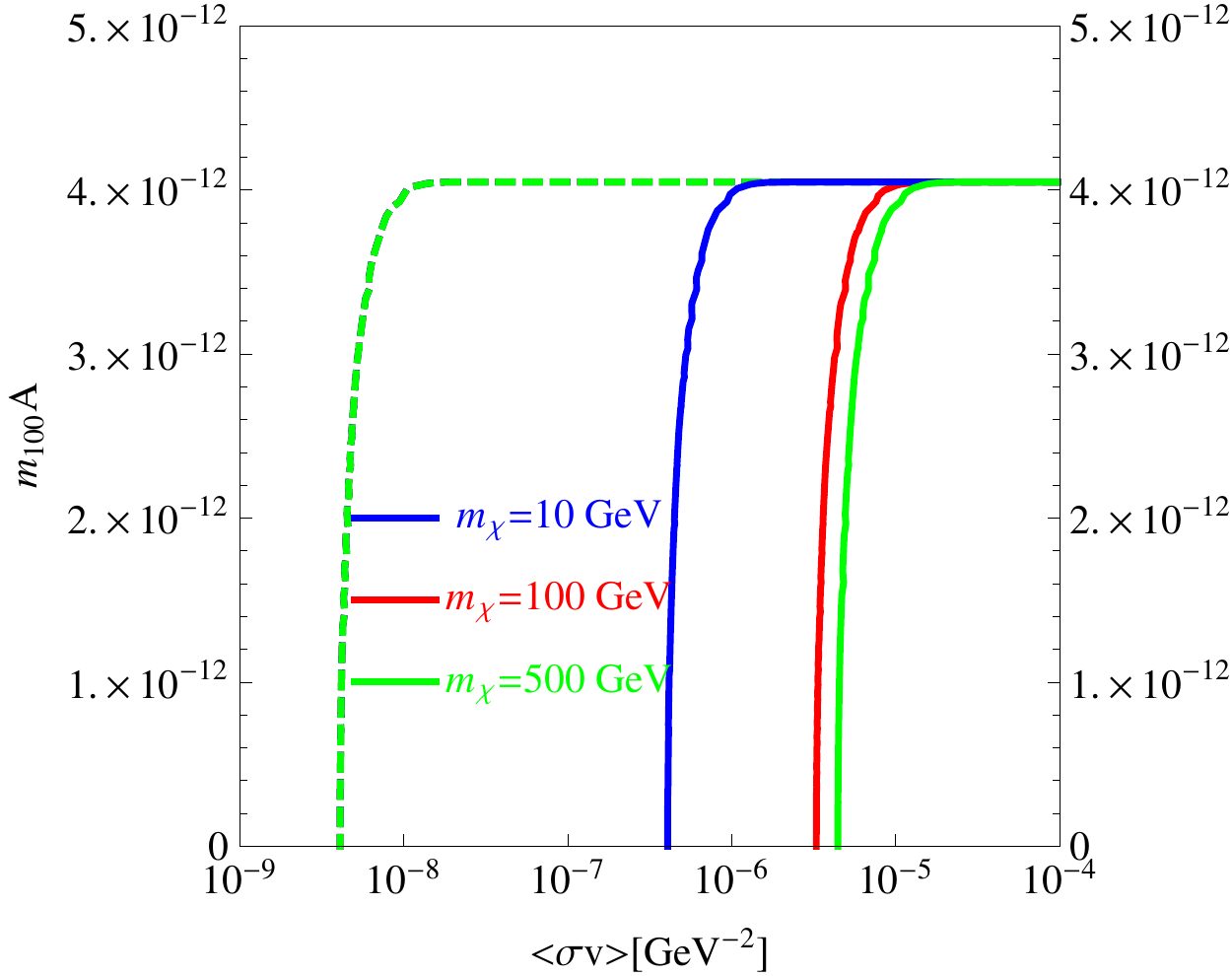}
\caption{Same as Fig.~\ref{fig:Mass-STD} but for \ref{fig:Mass}.a (left) kination   and  \ref{fig:Mass}.b. (right) the
 scalar-tensor model.  The solid contours correspond to a transition temperature of $T_{tr}= 5$ MeV and the dashed ones to $T_{tr}$ close to the standard cosmology freeze-out temperature.  In \ref{fig:Mass}.b the dashed lines are superimposed.}
\label{fig:Mass}
\end{center}
\end{figure}

The last case, in which the final $n_{\bar{\chi}}$ is smaller but not much smaller than ${n_{\chi}}$, happens  if $\bar{x}_{f o} > x_A$ but $\bar{x}_{f o} \simeq x_A$, i.e. if the decoupling  of $\bar\chi$ occurs shortly after the split between $n_{\chi}^{EQ}$ and  $n_{\bar{\chi}}^{EQ}$.  Given a particular asymmetry $A$, the  range of cross sections that satisfies this relationship is  narrow,  approximately within a factor of three. We choose $A$ such that $\chi$ and $\bar{\chi}$ constitute all of the DM, Eq.~\eqref{Omegas}.  In the standard cosmology the necessary annihilation cross section  is very close to  that of symmetric DM satisfying the same condition, and as soon as the cross section departs by a factor of a few from this value, the relic minority component density becomes exponentially small.  In the kination and scalar-tensor cosmologies, for a fixed transition temperature, the required values of the annihilation cross section in this third case can be much larger than the standard symmetric cross section (although in each case the  values must be in a reatively narrow range, within a factor of a few). This is illustrated in Fig.~\ref{new-fig:6}, which
presents the contours of $m_{100}A$ (recall $m_{100}=m_{\chi}/100$ GeV) as a function of the  $\chi\bar{\chi}$ annihilation cross section for which $\chi$ and $\bar{\chi}$ make up all of the DM (Eq.~\eqref{Omegas} holds).  As exemplified  in Figs.~\ref{fig:Mass-STD} and \ref{fig:Mass}, the $m_{100}A$ contours are almost independent $m_\chi$ in the standard cosmology and  if  $m_\chi > 100$ GeV in the other two.  The contours of Fig.~\ref{new-fig:6} do not apply  if  $m_\chi \simeq 10 $ GeV and $T_{tr}$ is close to  5 MeV (see Fig.~\ref{fig:Mass}).  We use $\Omega_{DM}=(\rho_{\chi}+\rho_{\bar{\chi}})/\rho_{c}$, with $\Omega_{DM}=0.11  h^{-2}$, $\rho_c=1.05375 \times 10 ^{-5}h^2\,{\rm cm}^{-3}$,  $\rho_{\chi}=m_{\chi}Y_{\chi}s_0$ and  $\rho_{\bar{\chi}}=m_{\chi}Y_{\bar{\chi}}s_0$ where $s_0=2889.2\, {\rm cm}^{-3}$ is  the present entropy density of the Universe \cite{Beringer:2012}. With these values of the various parameters,  the $m_{100} A$ contours satisfy the condition 
\begin{equation}
m_{100}(Y_{\chi}+Y_{\bar{\chi}})=4.01 \times 10^{-12}.
\label{A-condition}
\end{equation}
   For the standard cosmology, there is only one contour that satisfies Eq.~\eqref{A-condition} (the left most contour, shown in blue). In kination and scalar tensor models the contour satisfying the same condition depends on the transition temperature  $T_{tr}$, which in relevant models can take values between somewhat below 5 MeV and the standard cosmology freeze-out temperature. Thus, there is a range of contours which satisfy Eq.~\eqref{A-condition} for the kination and scalar-tensor models, as shown (dashed gray lines) in Fig. \ref{new-fig:6}.a and Fig. \ref{new-fig:6}.b respectively.   Values of $m_{100}A$ and $\langle \sigma_{\chi{\bar{\chi}}}  v \rangle$  above and to the left of the contours would lead to a relic density larger than the DM density.  Given a particular contour, values of $\langle \sigma_{\chi{\bar{\chi}}}  v \rangle$ and $m_{100} A$ below it would lead to a relic density of $\chi$ and ${\bar{\chi}}$ smaller than the DM density.

For each DM particle mass $m_\chi$ there is a maximum value of $A$, corresponding to  the horizontal part of each contour, which satisfies Eq.~\eqref{A-condition} with $Y_{\chi}+Y_{\bar{\chi}}=A$, which (since $Y_{\chi}-Y_{\bar{\chi}}=A$) means that $Y_{\chi}=A$ and $Y_{\bar{\chi}}$ is negligible. Thus the horizontal part each  contour, common to all of them, corresponds to the second case mentioned above, where $\bar{x}_{f o} \gg x_A$.  Only in the curved and vertical parts of the contours can $Y_{\bar{\chi}}$ be non-negligible with respect to $Y_{\chi}$. As the asymmetry $A$ becomes smaller, the contribution of  $\bar{\chi}$ to the relic density increases, and when $A$=0 we recover the symmetric case. Eq.~\eqref{A-condition} and the definition of $A$ imply that for values of $m_{100}A$ ranging from 0 to $3.3\times 10^{-12}$,  the relic density of $\bar{\chi}$ must be within a factor of 10 of the density of $\chi$.  The curved part of the contour corresponds to the third case mentioned above, where $\bar{x}_{f o}$ is larger but very similar to $x_A$, for which the asymmetry is large but the relic density of $\bar{\chi}$ is only a few orders of magnitude smaller than the density of $\chi$.

As already mentioned the  contours in Fig.~\ref{new-fig:6} are only slightly affected by changes in $m_\chi$ in the standard cosmology, and  if  $m_\chi > 100$ GeV also in kination and scalar-tensor models. Fig.~\ref{fig:Mass-STD} shows the  almost overlapping $m_{100} A$ contours for which $\chi$ and $\bar{\chi}$ constitute the whole to the DM in the standard cosmology for $m_\chi$ of 10 GeV (red), 100 GeV (green), and 500 GeV (blue). The analytic solutions predict this behavior.  For example, one can calculate the minimum annihilation cross section for a mass $m_{\chi}$, which occurs when $m_{100}A\rightarrow 0$.  For small $m_{100}A$, the exponential term in the analytic solutions (Eqs.~\eqref{Ybarsolstd} and \eqref{Ysolstd}) can be expanded as ${\rm exp}[A\lambda a/\bar{x}_{fo}]\simeq 1+A\lambda a/\bar{x}_{fo}$.  With this approximation, and substituting the analytic solutions into Eq.~\eqref{A-condition}, we see that $2m_{100}\bar{x}_{fo}/(\lambda a)=4.01\times 10^{-12}$.  Recalling that $\lambda = 3.0 \times 10^{21}m_{100}\, {\rm GeV}^2$ and using $\bar{x}_{fo}\simeq 20$, as in the symmetric case, we calculate a minimum cross section $a\simeq3.3\times 10^{-9}\, {\rm GeV}^{-2}$, independent of $m_{\chi}$.

Figs.~\ref{fig:Mass}.a and \ref{fig:Mass}.b show the same as Fig.~\ref{fig:Mass-STD} but for the kination and scalar-tensor models, respectively.  The solid contours correspond to models with a transition temperature of $T_{tr}=4$ MeV, and the dashed contours to $T_{tr}$ close to  the standard cosmology freeze-out temperature.  The figure shows that for $m_\chi=10$  GeV the $m_{100} A$ contours  change considerably with respect of those for $m_\chi >100$ GeV only  if $T_{tr}$ is low.

Notice that the value of $m_{100}A$ in the horizontal part of the  contours is unaffected by changes in $m_\chi$ in all three models, Figs.~\ref{fig:Mass-STD}   and \ref{fig:Mass}. Clearly, this must be so because there $Y_\chi =A$, and $Y_{\bar{\chi}}$ is negligible, thus the fixed DM density depends only on the product $m_\chi A$.
 \begin{figure}
\begin{center}
\includegraphics[width=6.5cm]{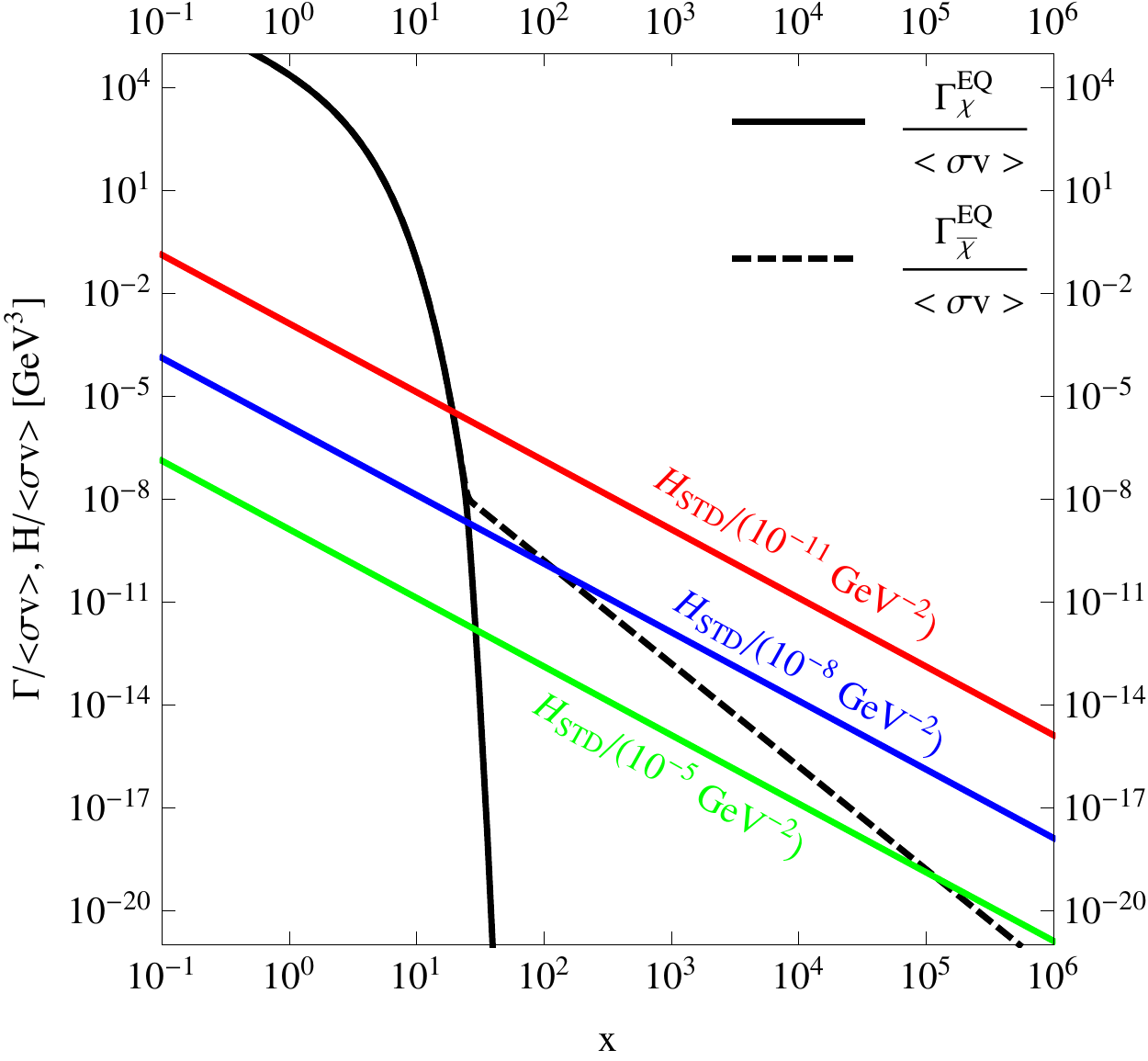}
\includegraphics[width=7.6cm]{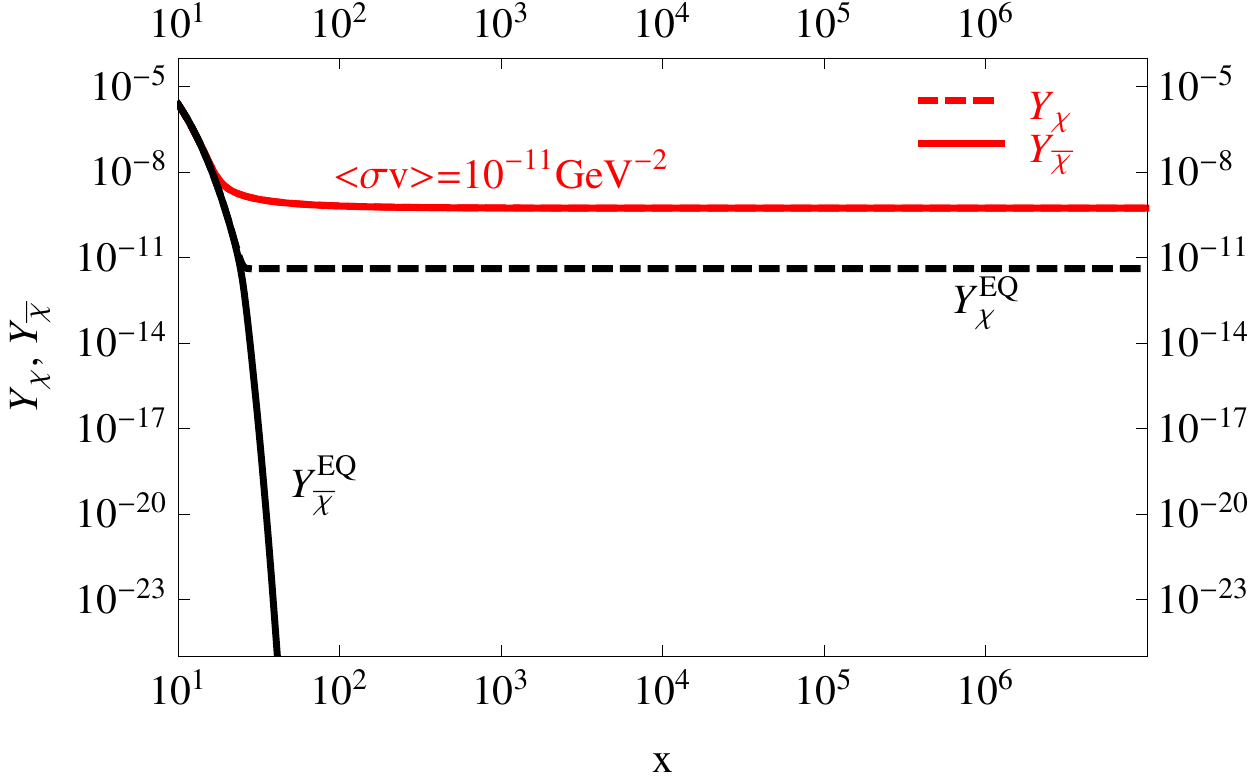}
\includegraphics[width=7.6cm]{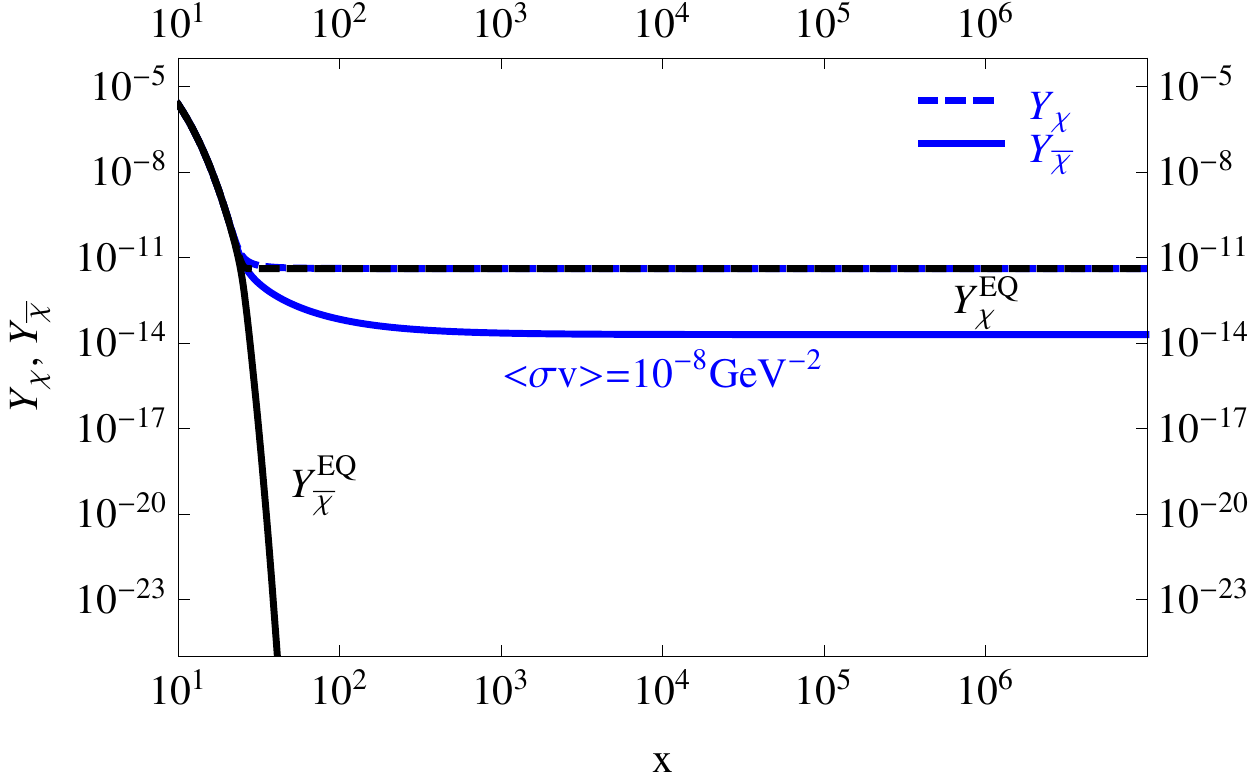}
\includegraphics[width=7.6cm]{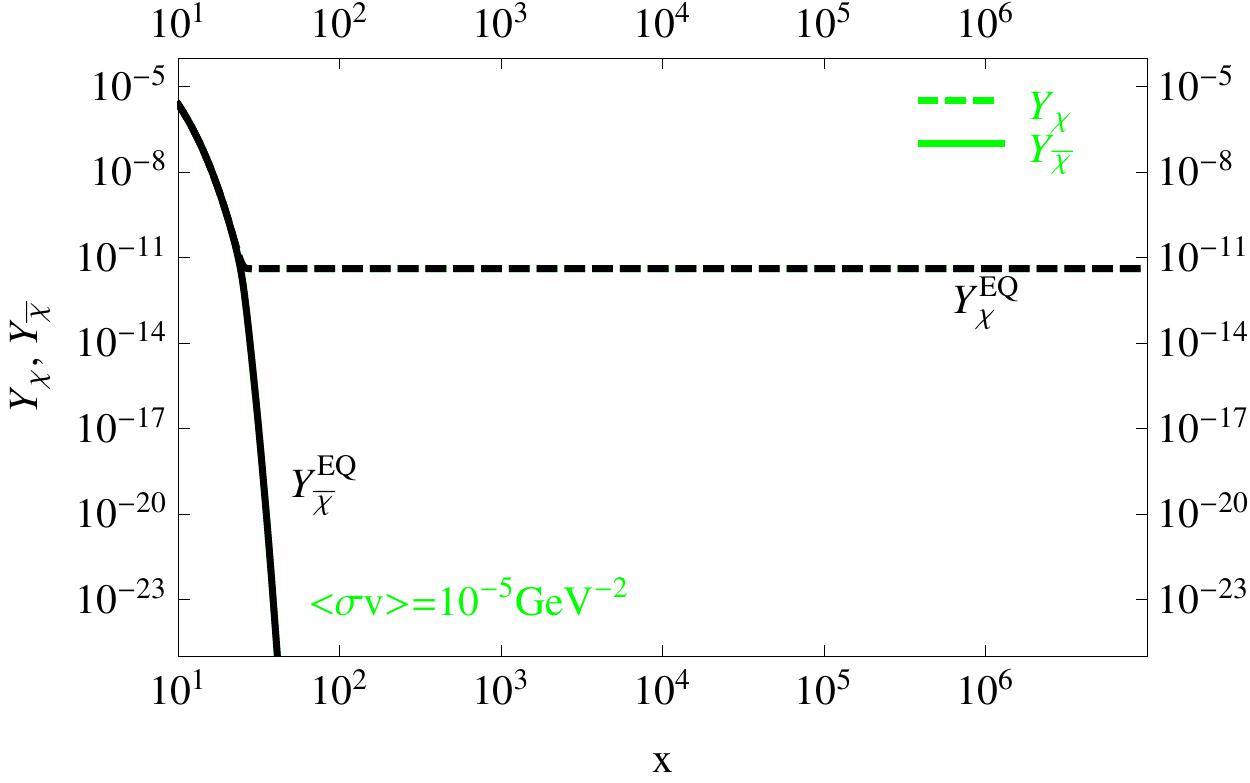}
\caption{In  \ref{fig:YH}.a (top left) the equilibrium annihilation rate over $\langle \sigma v \rangle =\langle \sigma_{\chi \bar{\chi}} v \rangle$ (black lines) is compared with the expansion rate in the standard cosmology over  $ \langle \sigma  v \rangle$, for three values of 
$\langle \sigma  v \rangle$, $1\times 10^{-11}{\rm GeV}^{-2}$ (top, red, line),  $1\times 10^{-8}{\rm GeV}^{-2}$ (middle, blue, line) and $1\times 10^{-5}{\rm GeV}^{-2}$ (lower, green, line).  \ref{fig:YH}.b (top right),  \ref{fig:YH}.c (bottom left)  and  \ref{fig:YH}.d (bottom right) show the  evolution of $Y_{\chi}$ (dashed colored line) and $Y_{\bar{\chi}}$ (solid colored line) compared to their equilibrium values (dashed and solid black lines) for the three values of $ \langle \sigma_{\chi{\bar{\chi}}}  v \rangle$ just mentioned, respectively, and $A=4.05\times 10^{-12}$.  In  \ref{fig:YH}.b $Y_{\chi}\simeq Y_{\bar{\chi}}$, so the solid and dashed red lines are superposed.  In \ref{fig:YH}.c  $Y_\chi$ (dashed blue line) closely follows equilibrium, so the dashed black and blue lines are superposed, and $Y_{\bar{\chi}}$ (solid blue line)  results less than 3 orders of magnitude smaller than $Y_\chi$.  In \ref{fig:YH}.d both $\chi$ and $\bar{\chi}$ closely follow their equilibrium abundances, so the black and green lines are superposed.}
\label{fig:YH}
\end{center}
\end{figure}

\begin{figure}
\begin{center}
\includegraphics[width=6.5cm]{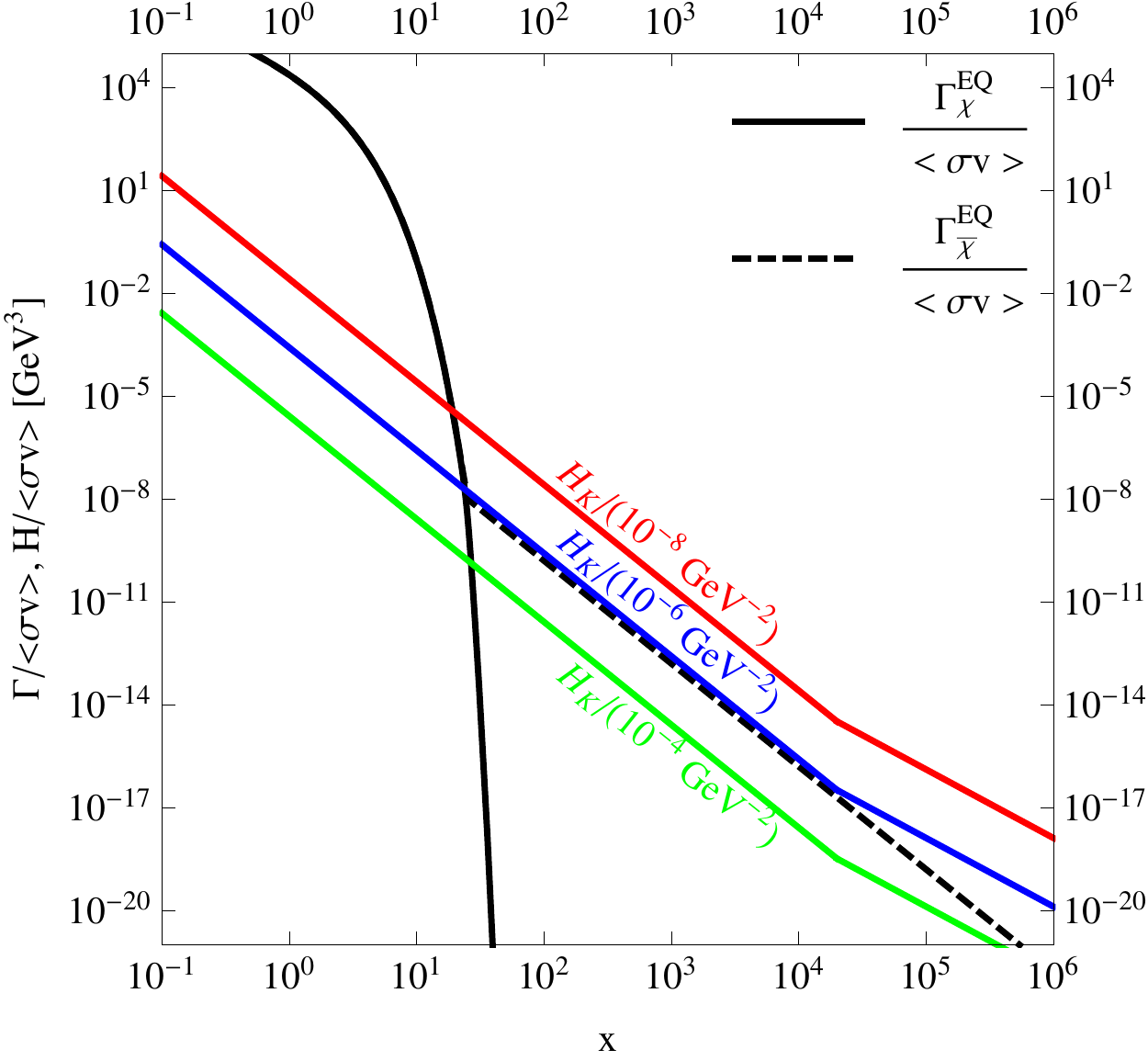}
\includegraphics[width=7.6cm]{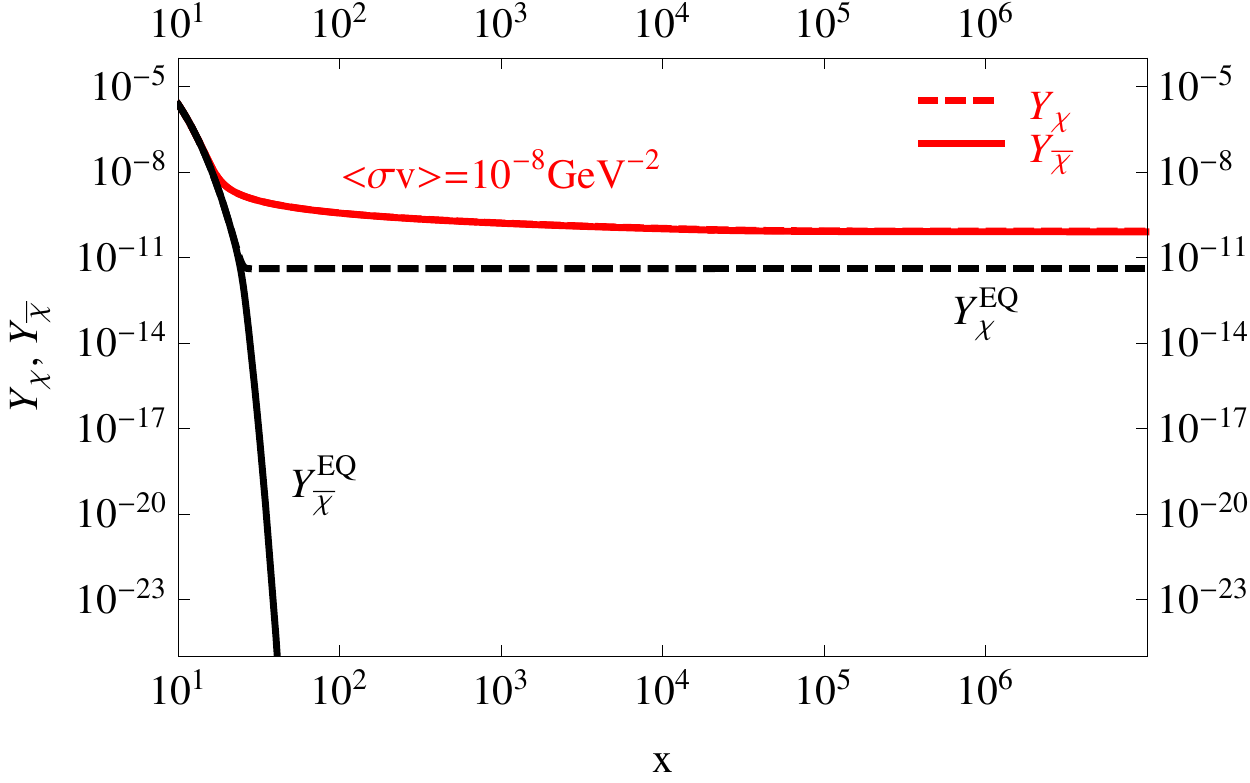}
\includegraphics[width=7.6cm]{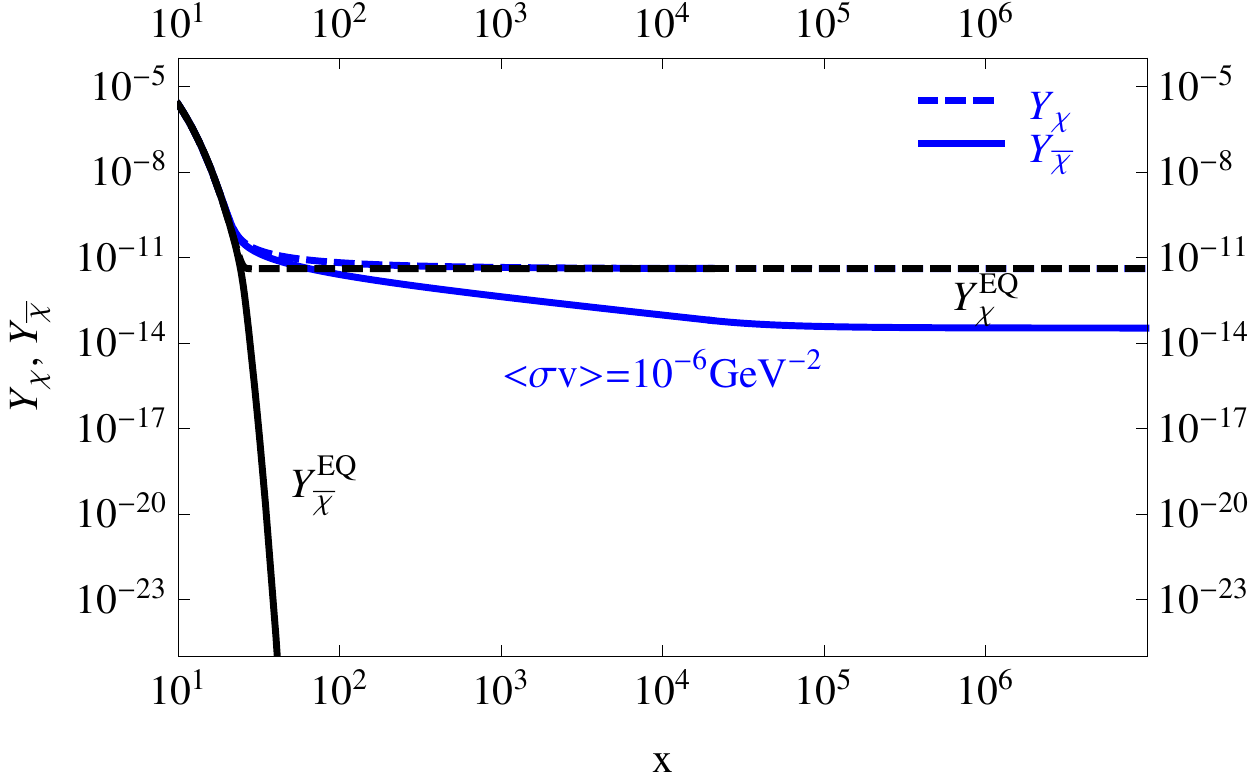}
\includegraphics[width=7.6cm]{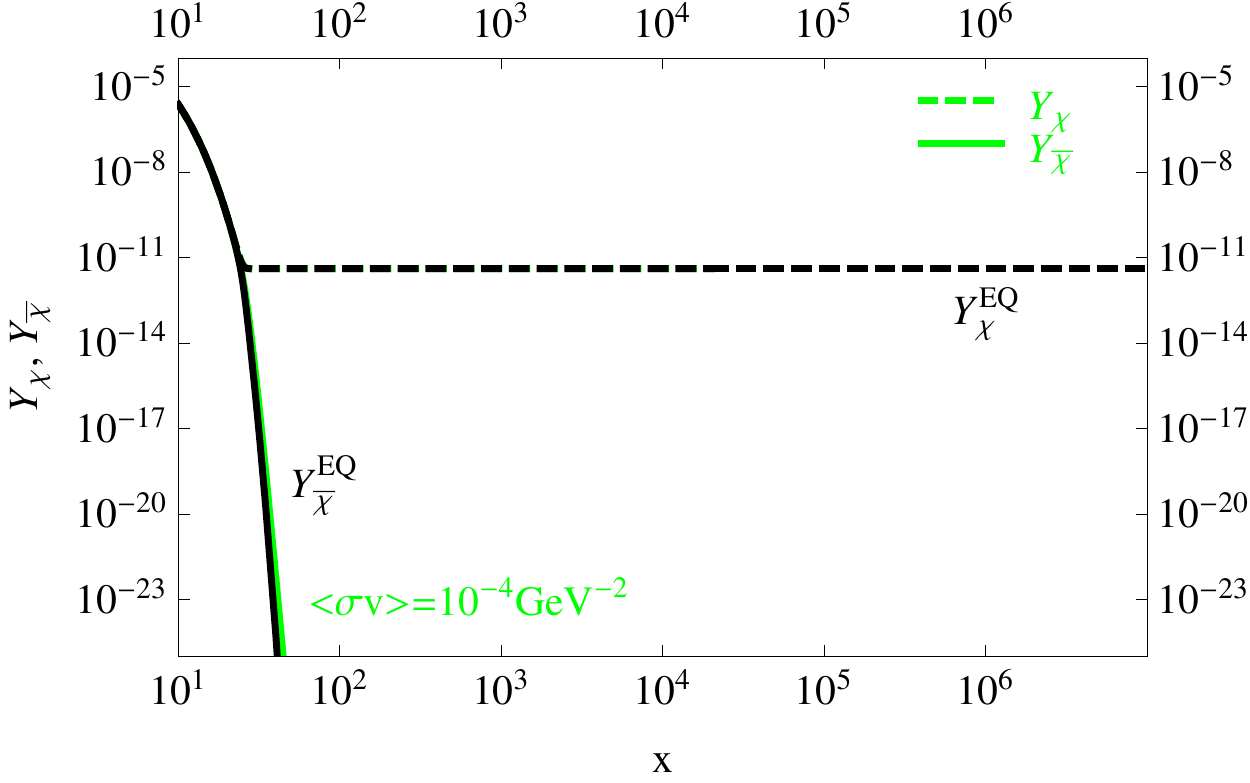}
\caption{Same as in Fig.~\ref{fig:YH} but for kination models and $\langle \sigma  v \rangle$ equal to $1\times 10^{-8}{\rm GeV}^{-2}$, $1\times 10^{-6}{\rm GeV}^{-2}$ and $1\times 10^{-4}{\rm GeV}^{-2}$, and $T_{tr}=5\,{\rm MeV}$. In Fig. \ref{fig:YHK}.d, $Y_{\bar{\chi}}$ is slightly greater than $Y_{\bar{\chi}}^{EQ}$.}
\label{fig:YHK}
\end{center}
\end{figure}

\begin{figure}
\begin{center}
\includegraphics[width=6.5cm]{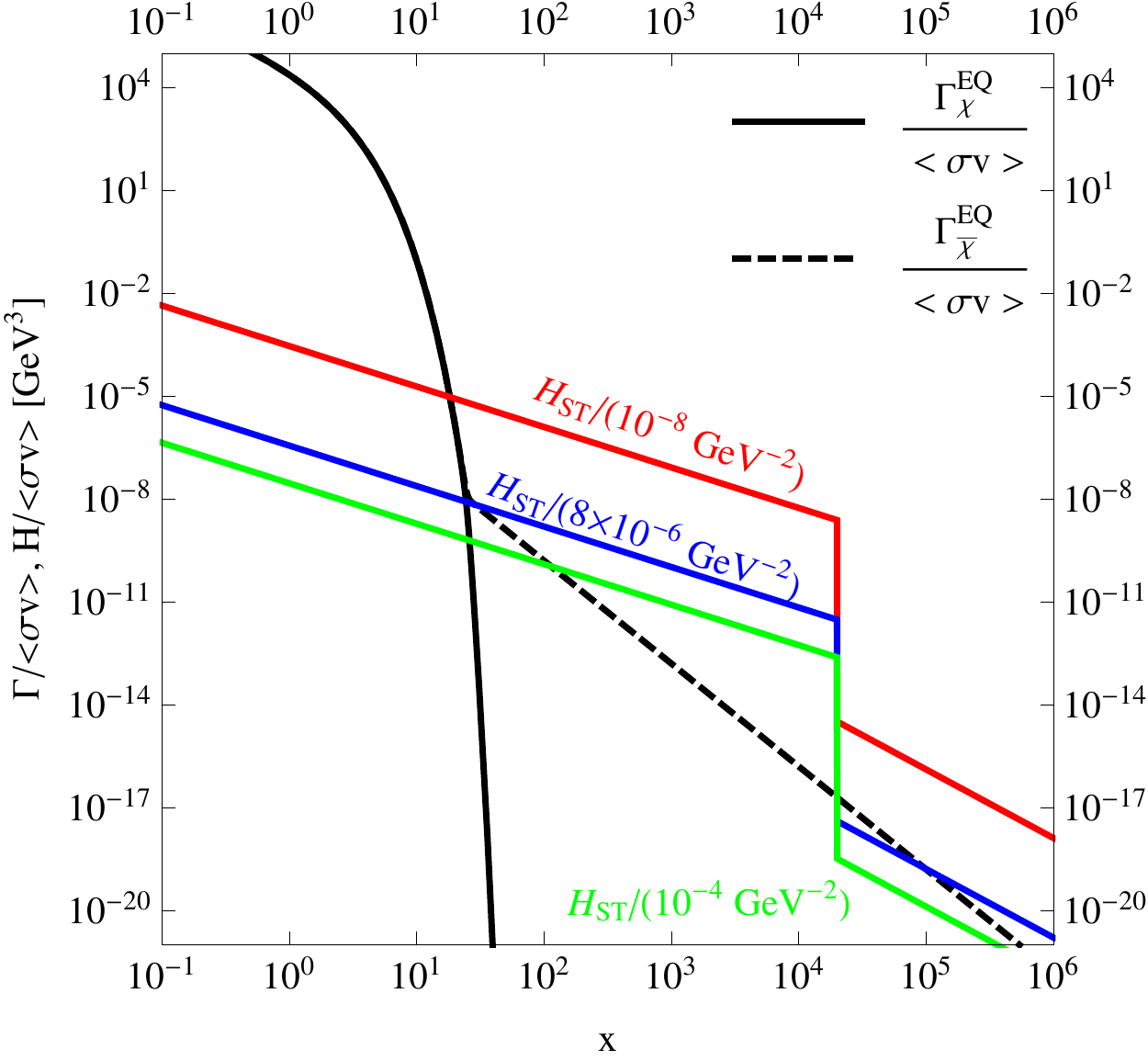}
\includegraphics[width=7.6cm]{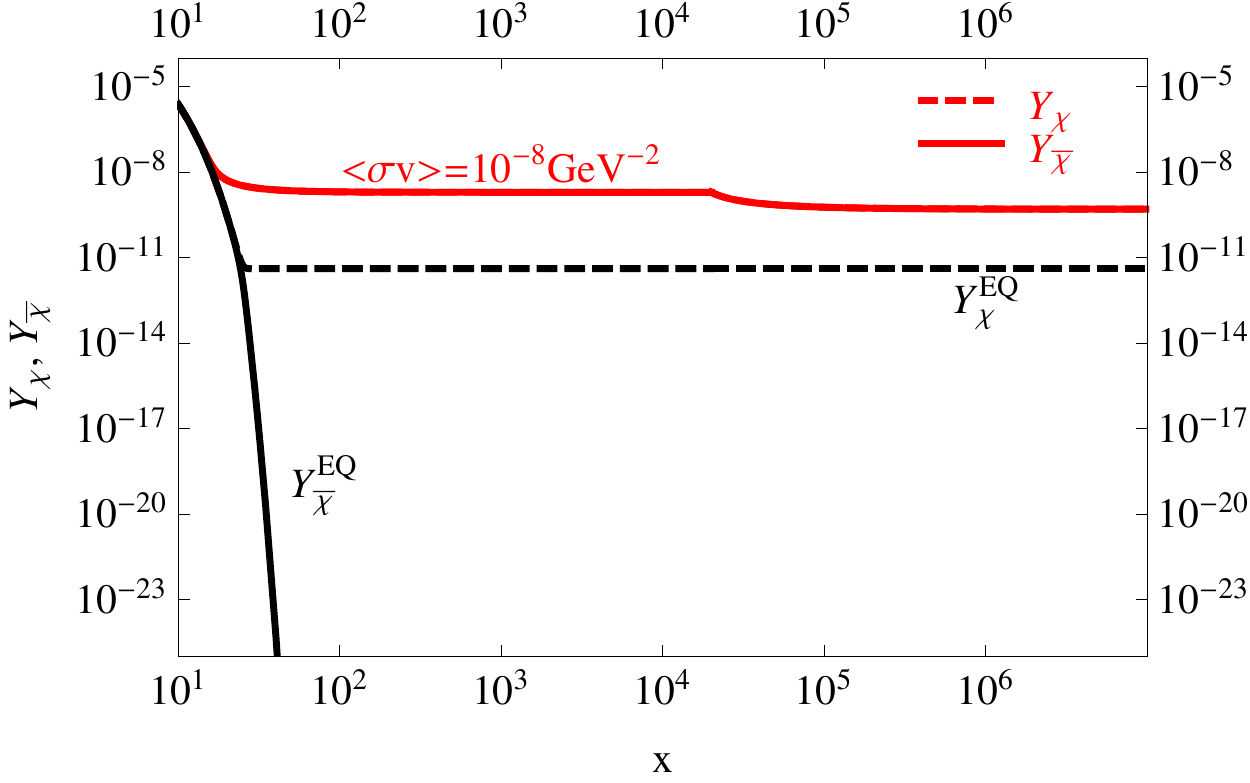}
\includegraphics[width=7.6cm]{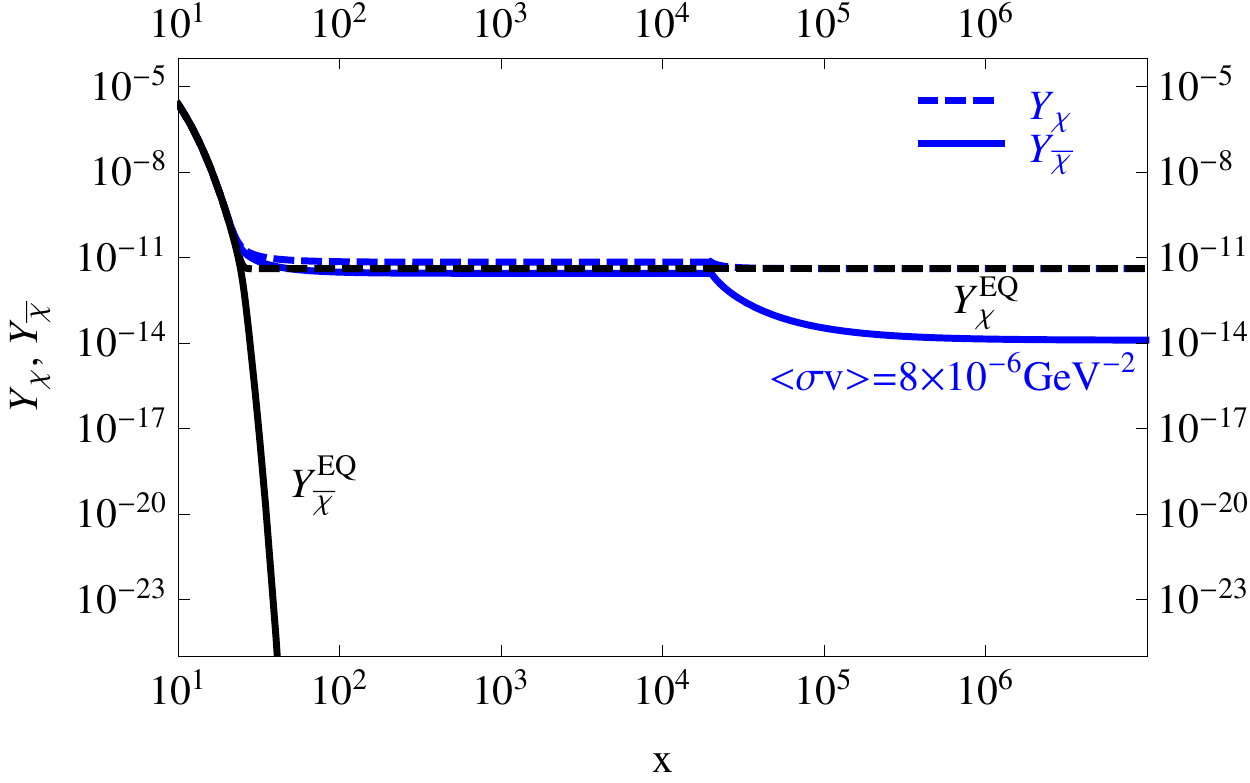}
\includegraphics[width=7.6cm]{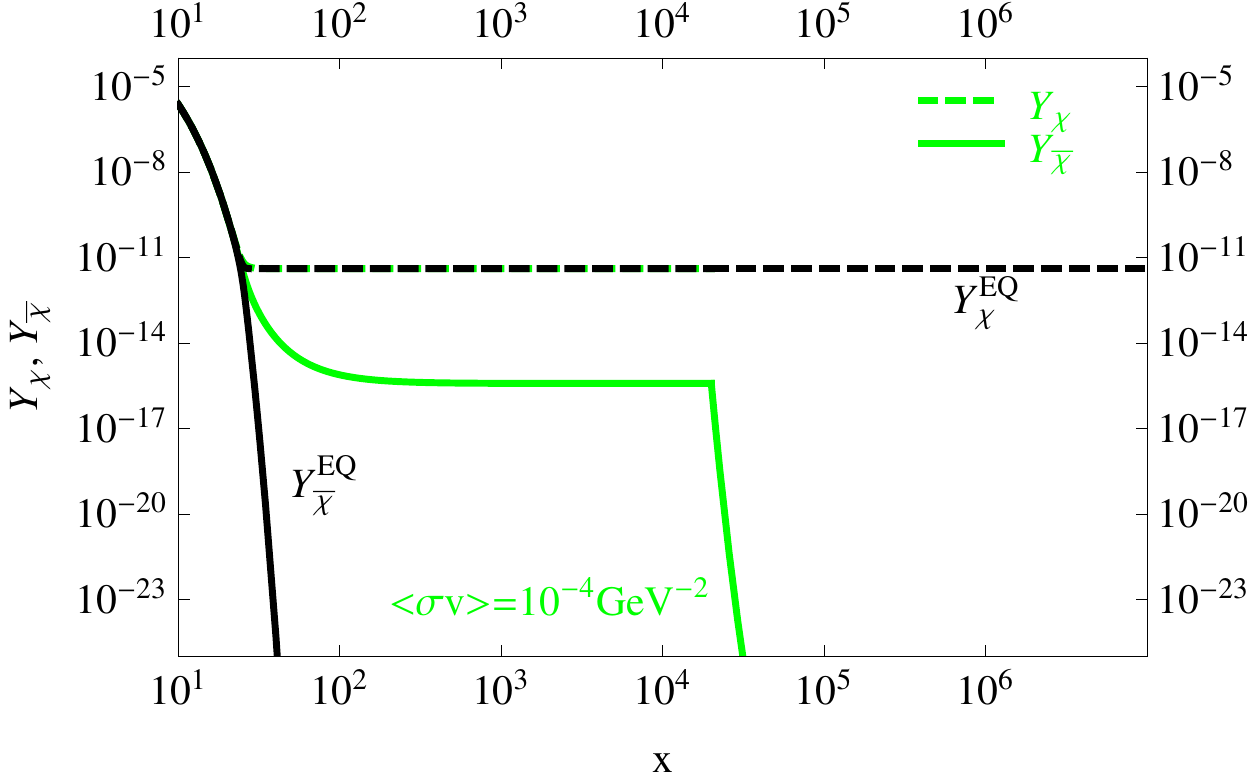}
\caption{Same as in Fig.~\ref{fig:YH} but for scalar tensor models and $\langle \sigma  v \rangle$ equal to $1\times 10^{-8}{\rm GeV}^{-2}$,  $8\times 10^{-6}{\rm GeV}^{-2}$, and  $1\times 10^{-4}{\rm GeV}^{-2}$, and $T_{tr}=5\,{\rm MeV}$. The effect or the re-annihilation phase is clearly seen. }
\label{fig:YHST}
\end{center}
\end{figure}

We mentioned three cases above in which either  $\bar{x}_{fo}< x_A$, $\bar{x}_{fo}\gg x_A$ or $\bar{x}_{fo}$ is larger but very close to $x_A$, so that the minority
relic number density $n_{\bar{\chi}}$ is either equal to,  negligible or just a few orders of magnitude smaller than the majority relic density $n_\chi$, respectively. Figs.~\ref{fig:YH}, \ref{fig:YHK} and \ref{fig:YHST} demonstrate these three cases in the standard cosmology, kination and scalar-tensor models respectively. In all three figures the bottom left panel shows the regime we are interested in, where the minority component relic density $n_{\bar{\chi}}$ is only a few orders of magnitude  smaller than the relic density of the majority DM component $n_\chi$. 


\section{Comparison of analytic and numerical relic abundance solutions}

\begin{figure}
\begin{center}
\includegraphics[width=7.6cm]{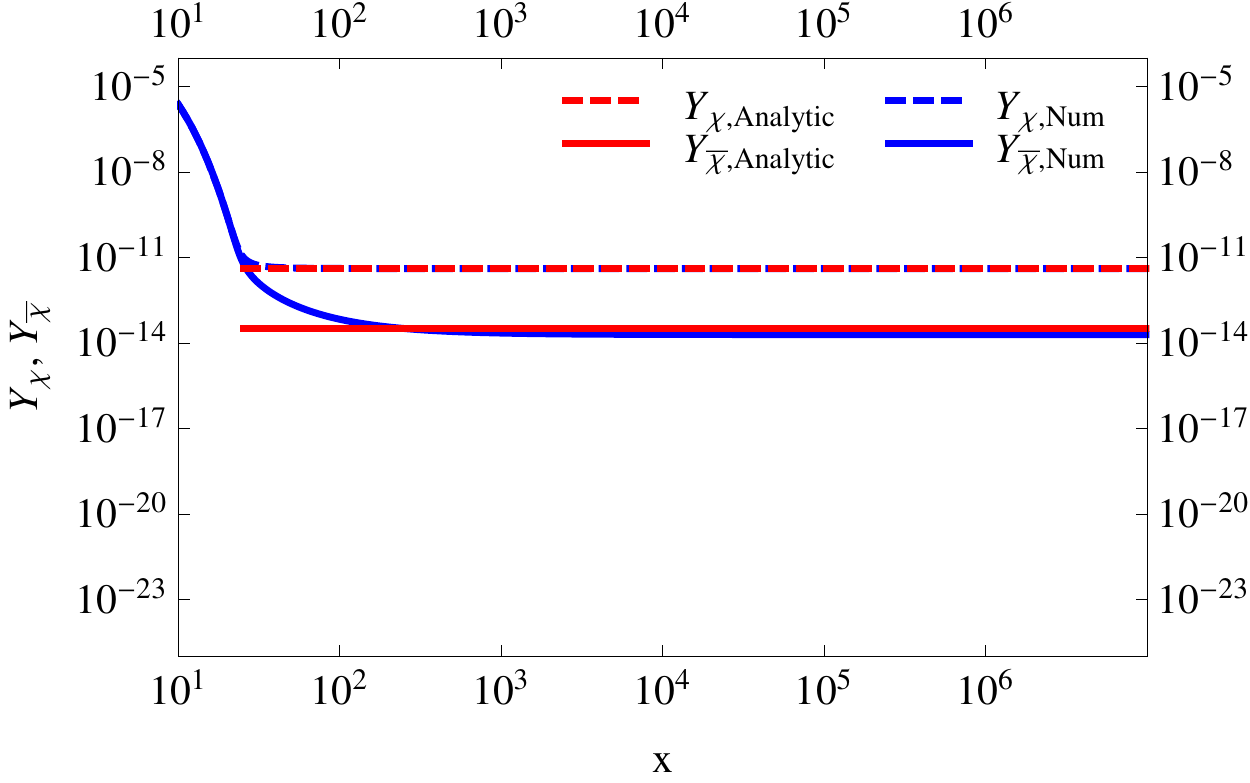}
\caption{Comparison of numerical (blue) solutions for $Y_{\chi}$ (dashed lines) and $Y_{\bar{\chi}}$ (solid lines) as a function of $x$ and analytical (red) solutions for $Y_{\chi}(x\rightarrow \infty)$ and $Y_{\bar{\chi}}(x\rightarrow \infty)$ in the standard cosmology. Here $m_{\chi}=$100 GeV, $A=4.01\times 10^{-12}$  and $\langle \sigma v \rangle = \langle \sigma_{\chi{\bar{\chi}}}  v \rangle= 1 \times 10^{-8} {\rm GeV}^{-2}$.  The analytical solutions are drawn only for $x$ larger than  $\bar{x}_{f o}$, which is larger but very close to $x_A$.}
\label{fig:Compare-STD}
\end{center}
\end{figure}

\begin{figure}
\begin{center}
\includegraphics[width=7.6cm]{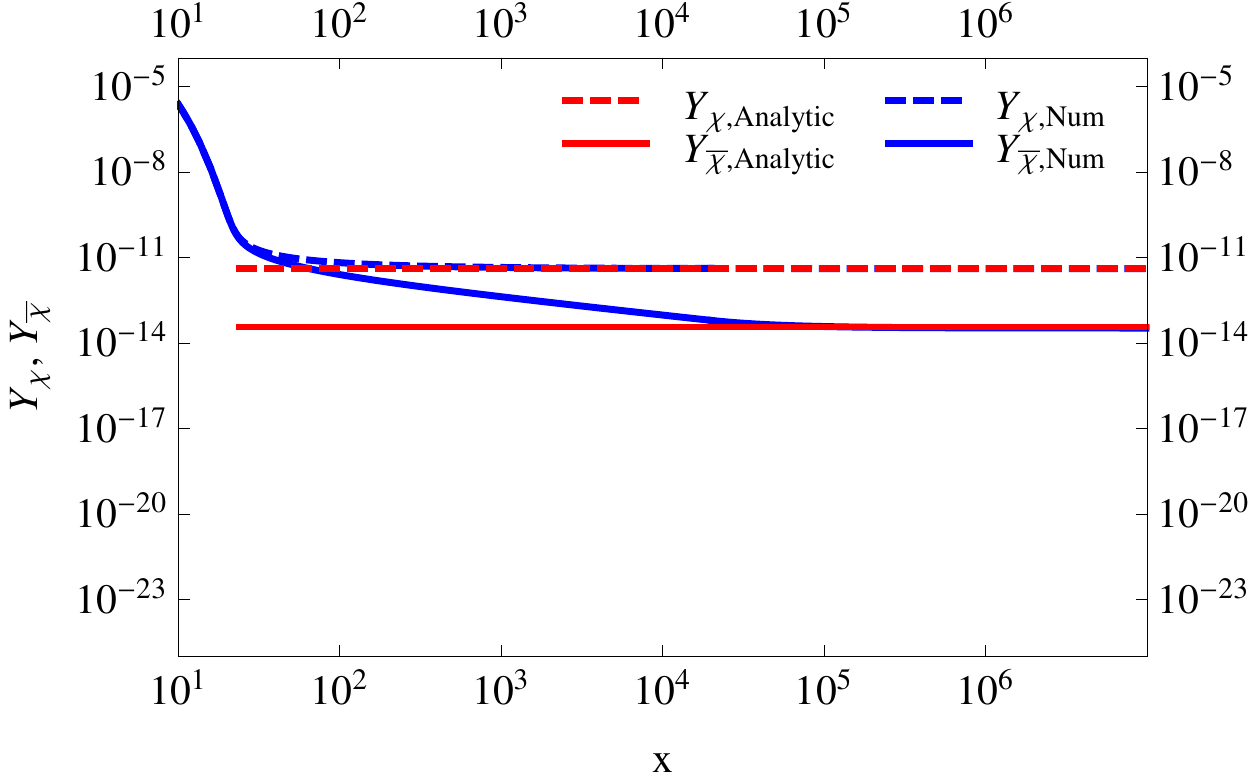}
\includegraphics[width=7.6cm]{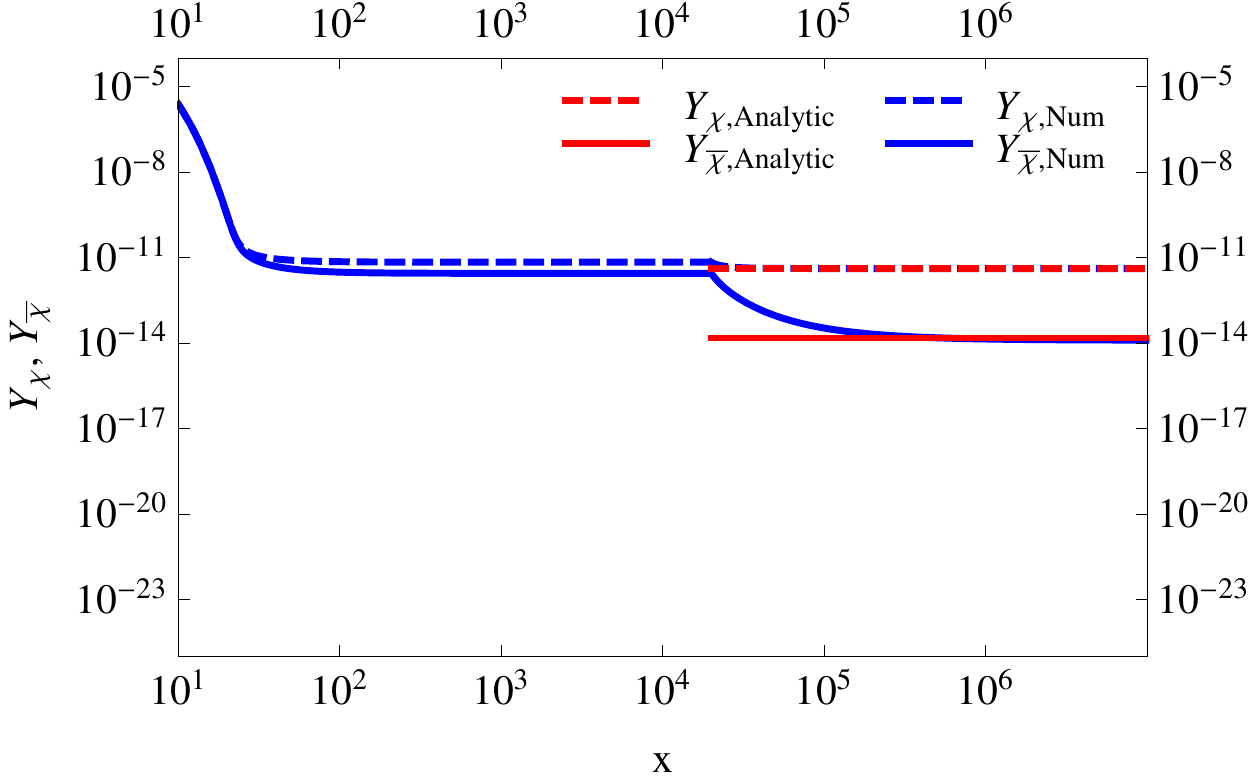}
\caption{Same as Fig.~\ref{fig:Compare-STD} but for  \ref{fig:Compare}.a (left) kination  and $\langle \sigma v \rangle = \langle \sigma_{\chi{\bar{\chi}}}  v \rangle = 1 \times 10^{-6} {\rm GeV}^{-2}$ and  \ref{fig:Compare}.b (left) scalar-tensor cosmology and $\langle \sigma  v \rangle = 8 \times 10^{-6} {\rm GeV}^{-2}$. In both cases $T_{tr}$=5 MeV.}
\label{fig:Compare}
\end{center}
\end{figure}

Figs.~\ref {fig:Compare-STD} and \ref{fig:Compare} show that the analytic solutions we found above agree very well
with the numerical solutions in all three cosmological models we examined. In  the examples presented, both solutions agree within a factor of two. In both figures $m_{\chi}=$100 GeV and $A=4.01\times 10^{-12}$  but the cross section for each particular cosmological model is chosen so that $\bar{x}_{f o}$ is not much larger than $x_A$,  because we wanted to show examples where the relic density  of the minority component is only a few orders of magnitude smaller than that of the majority component (see the figure captions). 

We have ascertained that the analytical solutions for the relic abundances  $Y_\chi$ and $Y_{\bar{\chi}}$ after $\bar{\chi}$ freeze-out reproduce the numerical solution to within a factor of four while  $Y_{\bar{\chi}}/ Y_\chi > 10 ^{-10}$ (smaller ratios are completely irrelevant). As this ratio increases, the agreement is better, within a factor of two.

\section{Present asymmetric DM annihilation rates}

Given a particular value of the  $\chi {\bar{\chi}}$ annihilation cross section and almost identical values of the asymmetry, very different  minority over majority asymmetric DM density ratios may result in the three pre-BBN cosmologies we explored, if the chemical decoupling  happens during these different cosmological phases.   Thus, if the annihilation cross section of the DM candidate is known, the annihilation rate at present, if detectable,  could be used to test the Universe before BBN, an epoch from which we do not  yet have  any data. In particular, because the annihilation cross section of asymmetric DM can be larger than that of symmetric DM in the standard cosmology, the 
decrease in the minority component relic density may be compensated or even overcompensated by the increase in annihilation cross section so that the annihilation rate at present of asymmetric DM, contrary to general  belief (see e.g. Ref.~\cite{Frandsen:2010zz}), could be non-negligible and even larger than that of symmetric DM in the standard cosmology.  There are experimental upper bounds on how large the annihilation cross section of asymmetric DM could be, but they are model dependent (see e.g. Ref.~\cite{March-Russell}). Assuming a ``neutralino like" DM candidate in the standard cosmology (for which particle and anti-particle coincide) the standard annihilation rate at present can be written as
\begin{equation}
\Gamma^{STD}_{sym} =\frac{1}{2}\langle \sigma_{self} v \rangle ({\rho_{DM}^2}/{m_\chi^2}),
\label{gamma-STD}
\end{equation}
where the $1/2$ factor corresponds to identical particles  annihilating, 
$\rho_{DM}= \Omega_{DM} \rho_c$ is the DM energy density, $ \langle \sigma_{self} v \rangle$ is the velocity averaged self-annihilation cross section,  which is similar to the early Universe  thermally averaged s-wave dominated cross section $ \langle \sigma_{self} v \rangle=a$. For s-wave annihilation with $x_{fo}\simeq 20$, and assuming that the candidate accounts for the whole of the DM, this is approximately $ \langle \sigma_{self} v \rangle \simeq 1.8\times10^{-9}{\rm GeV}^{-2}$.  Note that $\langle \sigma_{self} v \rangle$ in the early Universe is equal to half of the annihilation cross section needed for non-identical particles with no asymmetry ($A=0$) to make up all of the DM.  Numerically, we find this cross section to be $\langle \sigma_{\chi \bar{\chi}} v \rangle\simeq 3.7\times 10^{-9}\, {\rm GeV}^{-2}$.

For asymmetric DM, the annihilation rate is given by $\Gamma_{asym}= \langle \sigma_{\chi\bar{\chi}} v \rangle ({\rho_{\chi}\rho_{\bar{\chi}}}/{m_{\chi}^2})$, and we have assumed all along that  $\rho_{\chi}+\rho_{\bar{\chi}}=\rho_{DM}$ and $n_{\chi}=\rho_{\chi}/m_{\chi}$, thus 
\begin{equation}
\Gamma_{\rm asym}= \langle \sigma_{\chi\bar{\chi}} v \rangle \frac{\rho_{DM}^2}{m_{\chi}^2}\frac{Y_{\chi}Y_{\bar{\chi}}}{(Y_{\chi}+Y_{\bar{\chi}})^2}.
\label{gamma}
\end{equation}
The ratio of the asymmetric DM annihilation rate to the standard symmetric annihilation rate is then
\begin{equation}
\frac{\Gamma_{asym}}{\Gamma^{STD}_{sym}}=\frac{\langle \sigma_{\chi\bar{\chi}} v \rangle}{\langle \sigma_{self} v \rangle}\frac{2Y_{\chi}Y_{\bar{\chi}}}{(Y_{\chi}+Y_{\bar{\chi}})^2}.
\label{gratio}
\end{equation}

\begin{figure}
\begin{center}$
\begin{array}{cc}
\includegraphics[width=7.5cm]{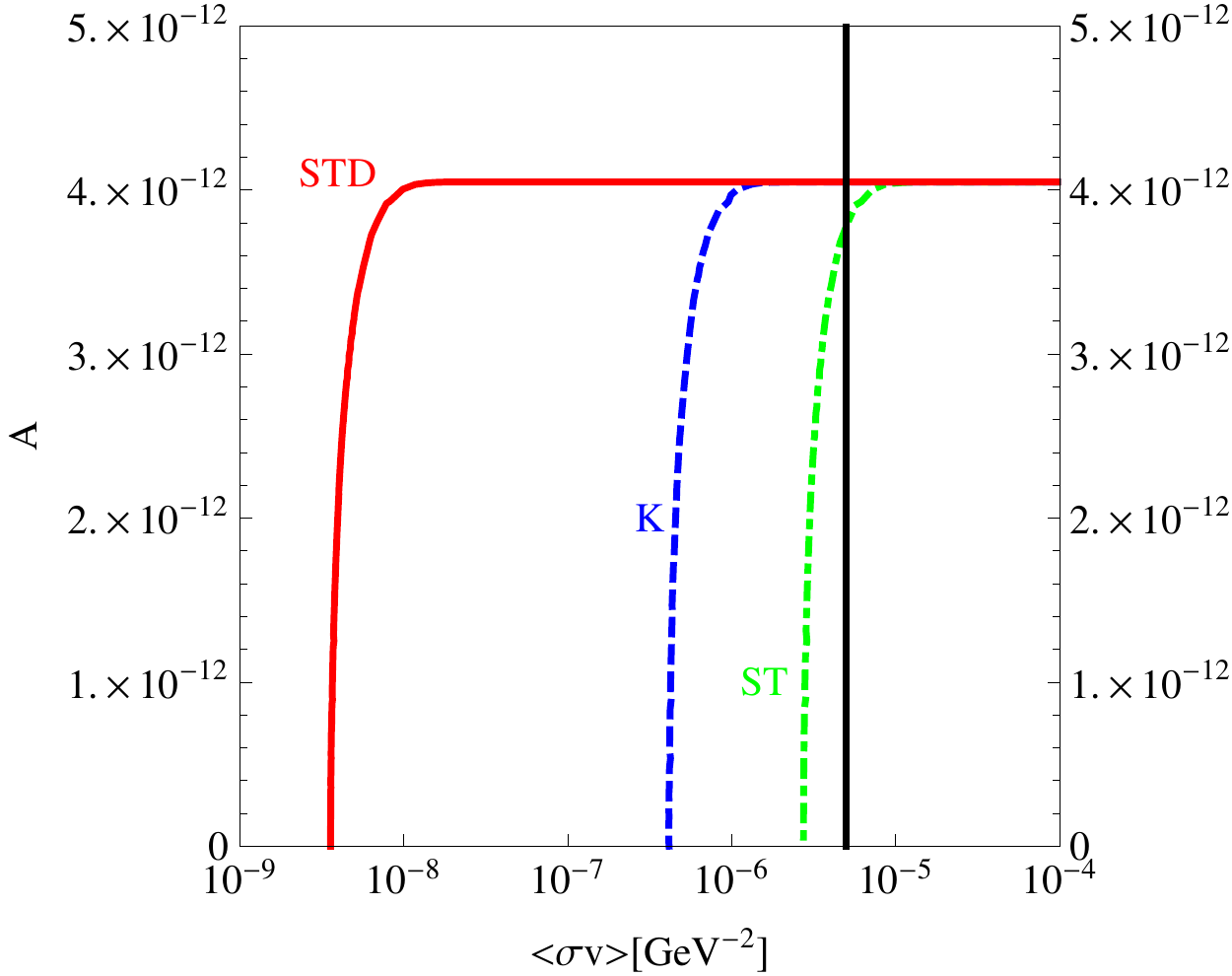}&
\includegraphics[width=7.5cm]{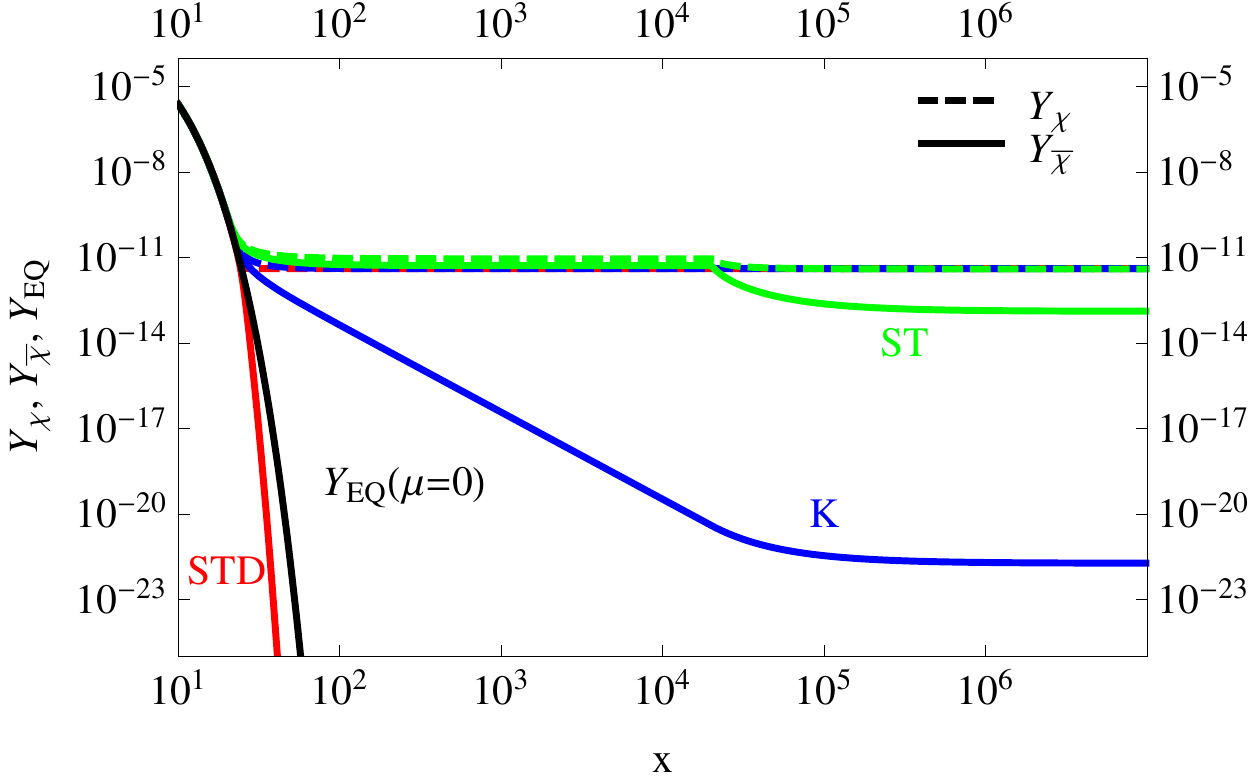}
\end{array}$
\caption{\ref{fig:3}.a (left) Contours of asymmetry $A$  for which $\chi$ and $\bar{\chi}$ make up all the DM  in  the standard cosmology (red), and in  kination  (blue) and scalar-tensor (green) models with $T_{tr}= $5 MeV, for $m_\chi=$100 GeV.  \ref{fig:3}.b (right) Evolution of $Y_\chi$ (dashed lines) and $Y_{\bar{\chi}}$ (solid lines) for $\langle \sigma v \rangle = \langle \sigma_{\chi{\bar{\chi}}}  v \rangle =5\times 10^{-6}\,{\rm GeV}^{-2}$ (indicated by the black vertical line in \ref{fig:3}.a)  in each of the three cosmological models.  For this cross section  $A=4.05\times 10^{-12}$ for the standard cosmology and kination models and $A=3.8\times 10^{-12}$ for the scalar-tensor model.  The evolution of $Y_{EQ}$ for symmetric DM ($A=$0) is shown for comparison.}
\label{fig:3}
\end{center}
\end{figure}

\begin{figure}
\begin{center}$
\begin{array}{cc}
\includegraphics[width=7.5cm]{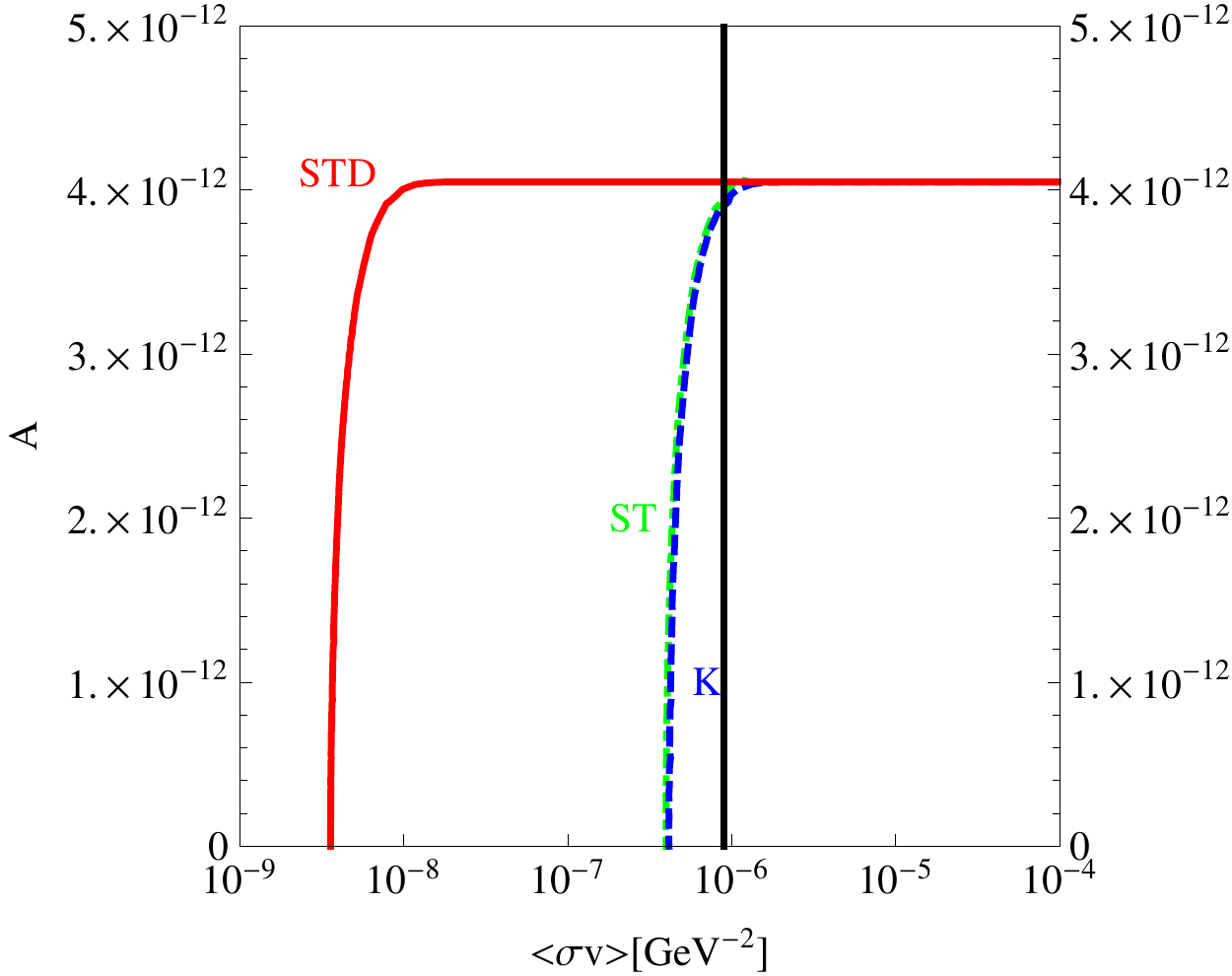}&
\includegraphics[width=7.5cm]{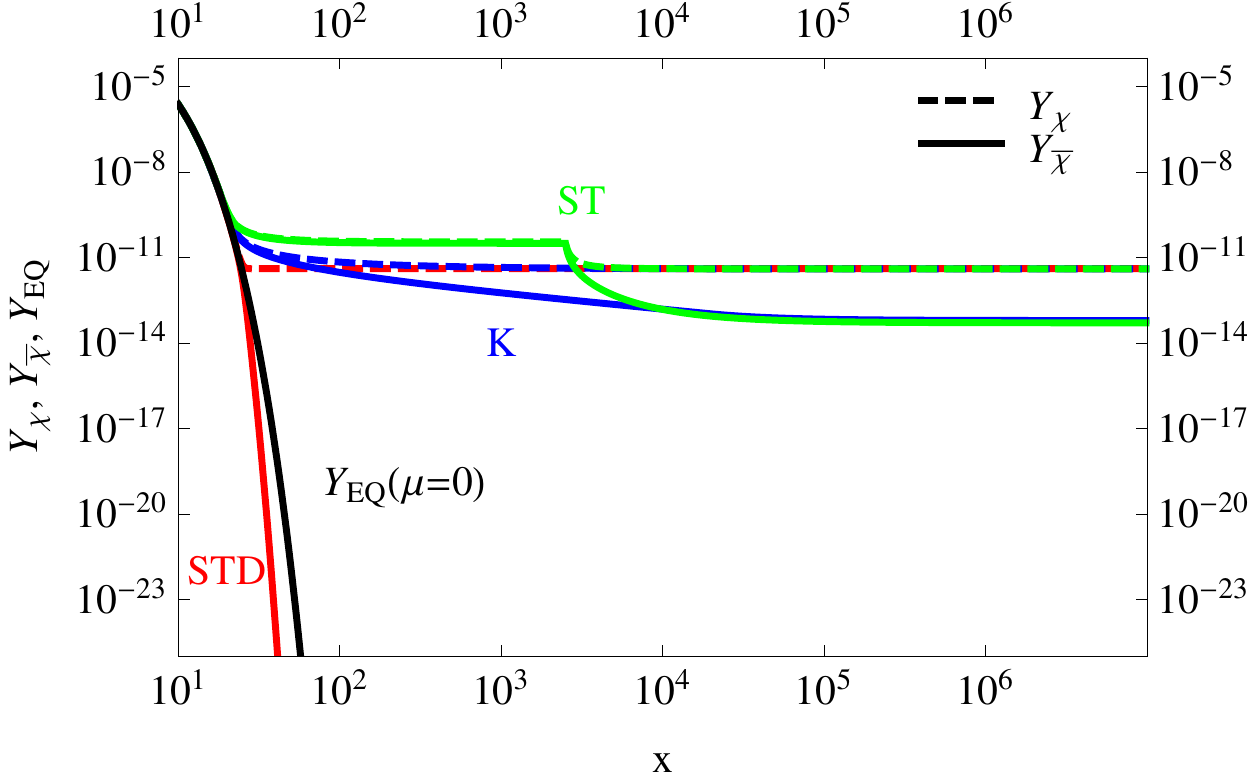}
\end{array}$
\caption{Same as in Fig.~\ref{fig:3} but with $T_{tr}=$ 40 MeV in the scalar-tensor model and  $\langle \sigma v \rangle = \langle \sigma_{\chi{\bar{\chi}}}  v \rangle =9\times 10^{-7}\,{\rm GeV}^{-2}$ for which $A=4.05\times 10^{-12}$ for the standard cosmology and  $A=3.9\times 10^{-12}$ for the kination and scalar-tensor models.}\label{fig:4}
\end{center}
\end{figure}

\begin{figure}
\begin{center}$
\begin{array}{cc}
\includegraphics[width=7.5cm]{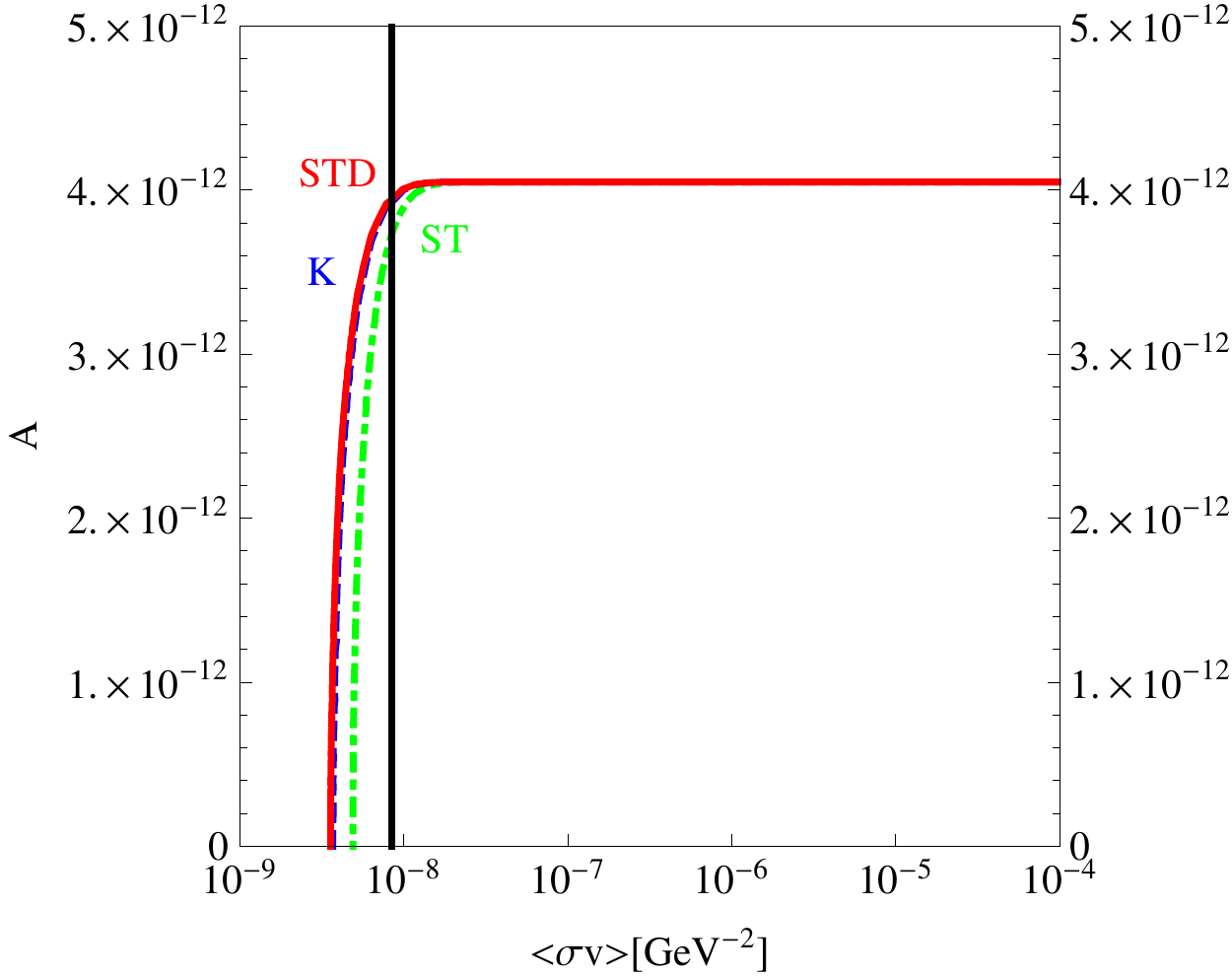}&
\includegraphics[width=7.5cm]{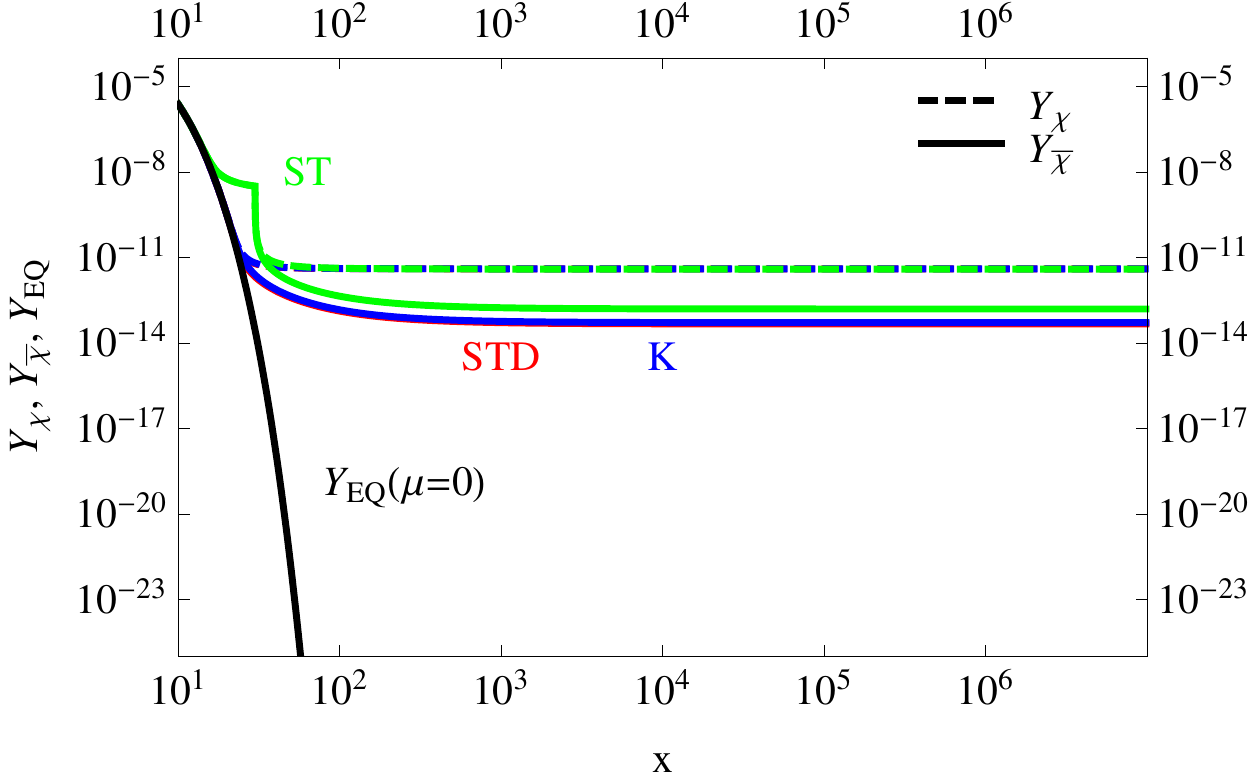}
\end{array}$
\caption{Same as in Fig.~\ref{fig:3} but with a high transition temperature $T_{tr}=$3.35 GeV in both the kination and scalar-tensor models and  $\langle \sigma v \rangle = \langle \sigma_{\chi{\bar{\chi}}}  v \rangle =8.5\times 10^{-9}\,{\rm GeV}^{-2}$ for which $A=3.95\times 10^{-12}$ for the standard cosmology and kination model, and $A=3.75\times 10^{-12}$ in the scalar-tensor model.  }
\label{fig:5}
\end{center}
\end{figure}

 In the standard cosmology, this ratio never exceeds one, but in the kination or  scalar-tensor pre-BBN cosmological models, it may exceed one for a range of  annihilation cross sections, as shown in Figs.~\ref{fig:3}, \ref{fig:4} and \ref{fig:5}.  This range of cross sections depends on the transition temperature $T_{tr}$ at which the cosmology becomes standard. 

 In Figs.~\ref{fig:3}, \ref{fig:4} and \ref{fig:5}  we provide examples  in which the asymmetric DM annihilation rate in a non-standard  pre-BBN cosmological model is greater than the  standard symmetric annihilation rate for $m_\chi=100\, {\rm GeV}$.  We have chosen particular  annihilation cross sections and transition temperatures in each case.  In  Fig.~\ref{fig:3}  the transition temperature is low, $T_{tr}$=5 MeV  and for $\langle \sigma_{\chi{\bar{\chi}}}  v \rangle=5\times 10^{-6}\,{\rm GeV}^{-2}$, i.e.  $\langle \sigma_{\chi\bar{\chi}} v \rangle/\langle \sigma_{self} v \rangle \sim 10^3$, for the scalar-tensor model we get $Y_{\chi}Y_{\bar{\chi}}/(Y_{\chi}+Y_{\bar{\chi}})^2\simeq 10^{-1}$, so the annihilation rate  is  $\sim 10^2$ times greater than the standard  symmetric DM annihilation rate.  For the kination and standard models the present annihilation rate is negligible in this example, $\sim 10^{-7}$  or smaller respectively than the standard symmetric rate. For kination,  $Y_{\chi}Y_{\bar{\chi}}/(Y_{\chi}+Y_{\bar{\chi}})^2\simeq 10^{-10}$, and this ratio is even smaller in the standard cosmology. Thus, in this example, the scalar-tensor model is easily distinguishable from the other two.

For smaller annihilation cross sections we can select kination and scalar-tensor models which would produce almost identical effects in the annihilation rate, as demonstrated in Fig. \ref{fig:4}.  In this figure, the transition temperatures, $T_{tr}$=5 MeV  for kination  and $T_{tr}=40\, {\rm MeV}$ for the scalar-tensor model, are chosen so that the contours of $A$  for which $\chi$ and $\bar{\chi}$ make up all the DM in both models are very similar.  For $\langle \sigma_{\chi{\bar{\chi}}}  v \rangle =9\times 10^{-7}\,{\rm GeV}^{-2}$, both models give a present asymmetric DM annihilation rate  $\sim 10$ times larger than the standard symmetric DM annihilation rate.  If such a rate were detected, we could see that the cosmology is non-standard, but could not distinguish between the kination and scalar-tensor models chosen here.

For asymmetric DM annihilation cross sections very close to the standard symmetric DM  annihilation cross section it is difficult to distinguish between any of the three pre-BBN cosmological models.  In order to prevent the Universe from being  overdense, the transition temperature of the non-standard cosmological models  must be set  very close to the standard cosmology freeze-out temperature. This means that $\chi$ and $\bar{\chi}$ only evolve in the non-standard cosmological phase for a very short time after freeze-out. In  Fig. \ref{fig:5}, $\langle \sigma_{\chi{\bar{\chi}}}  v \rangle =8.5\times 10^{-8}\,{\rm GeV}^{-2}$ and the transition temperature is high,  $T_{tr}=3.35\, {\rm GeV}$ for  both kination and scalar-tensor models.  All three models produce similar $\bar{\chi}$ relic densities, and the  present asymmetric DM annihilation rate is $\sim 10$ times smaller than the standard symmetric DM annihilation rate for all of them. Thus the annihilation rate could not help us distinguish between the different pre-BBN cosmological models.

These examples show that if the mass,  approximate particle-antiparticle asymmetry,  and annihilation cross section of  a DM candidate is known, then if the present annihilation rate is detectable,  one could possibly distinguish between these different pre-BBN cosmological models. A higher annihilation rate than predicted by the model of symmetric DM in the standard cosmology would imply the existence of a non-standard cosmology during pre-BBN epoch.  

\subsection{Fermi Space Telescope bounds on non-standard asymmetric DM models}

In non-standard pre-BBN cosmological models the annihilation rate of asymmetric DM can be so large as to already be rejected by present bounds. The Fermi-LAT collaboration has published upper bounds~\cite{Ackermann:2011}  on the annihilation cross section of a symmetric DM candidate $\langle \sigma_{self} v \rangle  <  \langle \sigma v \rangle_{\rm Fermi}$ as a function of its  mass, from a combined analysis of Milky Way satellites, assuming that it constitutes all of the DM and annihilates only into particular final states, which correspond to limits on the annihilation rate in Eq.~\eqref{gamma-STD},  $\Gamma^{STD}_{sym} < \Gamma_{\rm Fermi}$. Here we use the $\chi\bar{\chi}\rightarrow\mu\bar{\mu}$ and $\chi\bar{\chi}\rightarrow b\bar{b}$ modes, which impose the limit of $\langle \sigma v \rangle_{\rm Fermi}$ equal to $8.80 \times 10^{-25}$cm$^3/$s$= 7.33 \times 10^{-8}$ GeV$^{-2}$ and $5.99 \times 10^{-26}$cm$^3/$s$= 4.99 \times 10^{-9}$ GeV$^{-2}$ to 95\% C.L. for $m_\chi=100$ GeV, respectively. Thus, under the same assumptions used by the Fermi-LAT collaboration, for our models, the ratio
\begin{equation}
\frac{\Gamma_{asym}}{\Gamma_{\rm Fermi}}=\frac{\langle \sigma_{\chi\bar{\chi}} v \rangle}{\langle \sigma v \rangle_{\rm Fermi}}\frac{2Y_{\chi}Y_{\bar{\chi}}}{(Y_{\chi}+Y_{\bar{\chi}})^2}
\label{gfratio}
\end{equation}
must be less than 1.

 \begin{figure}
\begin{center}$
\begin{array}{cc}
\includegraphics[width=7.5cm]{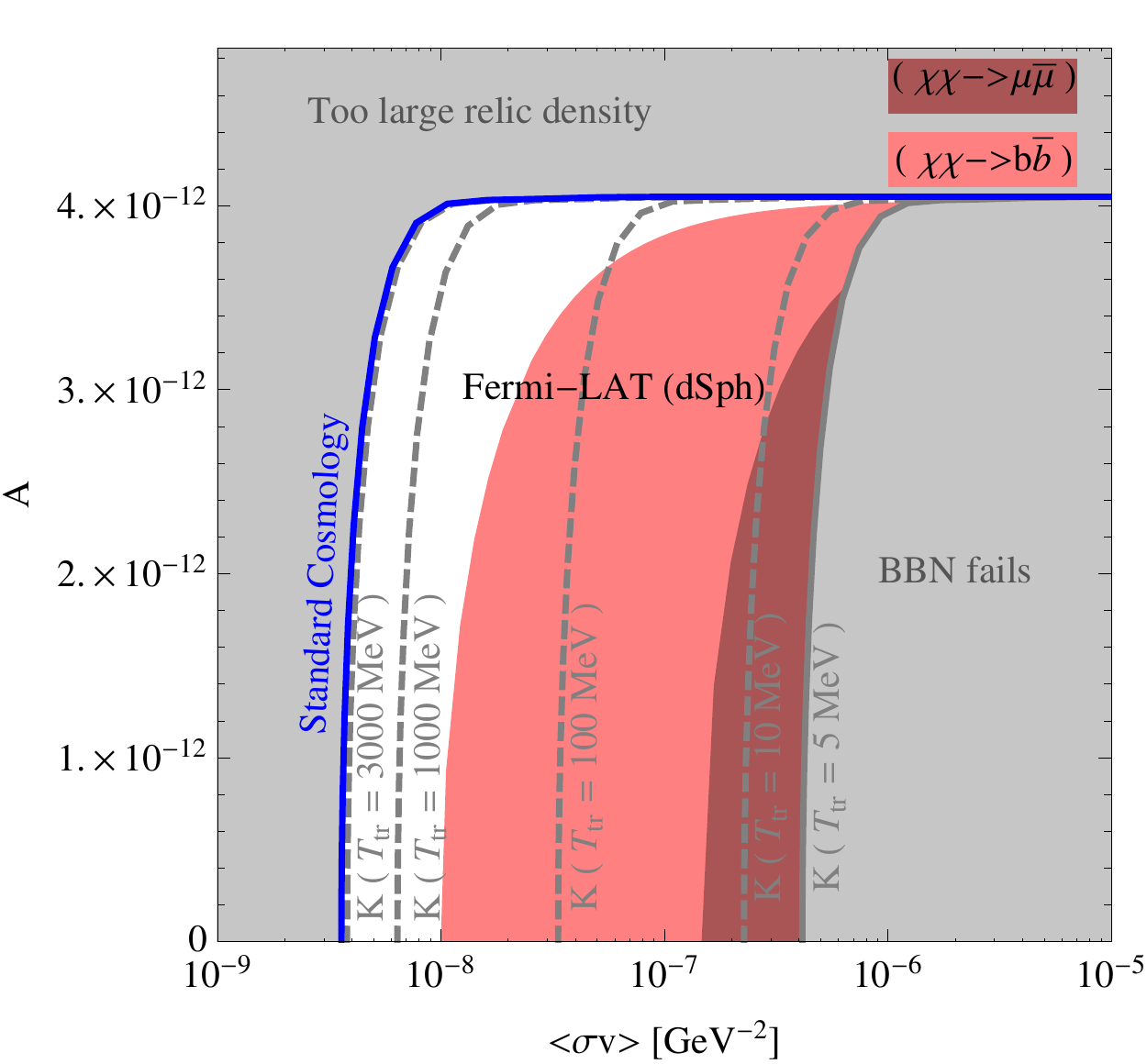}&
\includegraphics[width=7.5cm]{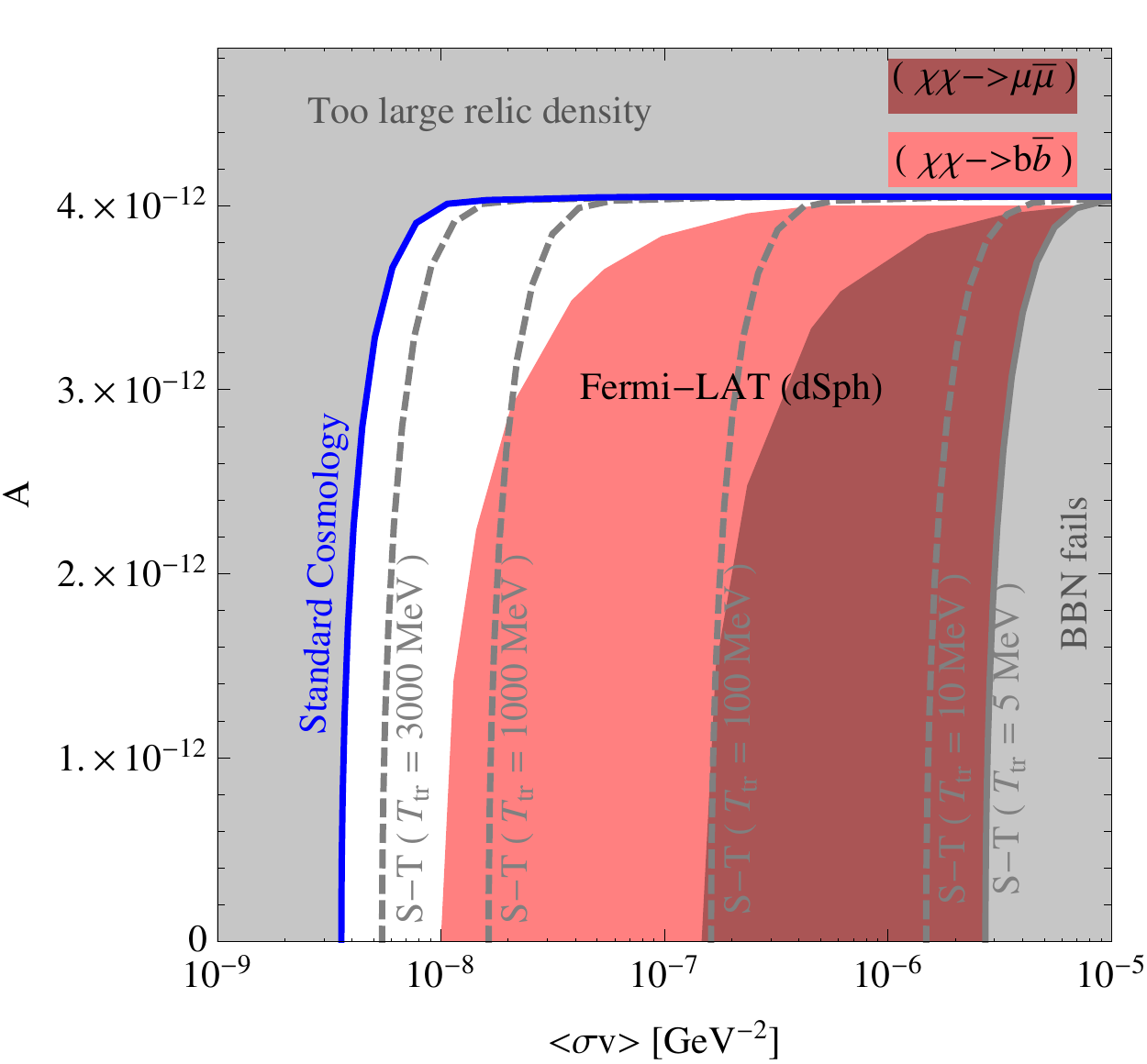}
\end{array}$
\caption{Regions of the DM asymmetry $A$ versus asymmetric DM  $\chi {\bar{\chi}}$ annihilation cross section $\langle \sigma v \rangle = \langle \sigma_{\chi{\bar{\chi}}}  v \rangle$  in  \ref{fig:6}.a (left) kination and \ref{fig:6}.b (right) scalar-tensor  pre-BBN cosmological models rejected by $\chi\bar{\chi}\rightarrow\mu\bar{\mu}$ (dark pink) and $\chi\bar{\chi}\rightarrow b\bar{b}$ (light pink) Fermi-LAT bounds~\cite{Ackermann:2011}  
 for $m_\chi = $100 GeV, under the same assumptions made by the Fermi-LAT collaboration.}
\label{fig:6}
\end{center}
\end{figure}

Fig. \ref{fig:6} shows (in pink) the  regions of the DM asymmetry $A$ versus asymmetric DM  $\chi {\bar{\chi}}$ annihilation cross section in  kination (Fig.~\ref{fig:6}.a) and  scalar-tensor (Fig.~\ref{fig:6}.b) pre-BBN cosmological models already rejected by Fermi-LAT bounds~\cite{Ackermann:2011} on the annihilation modes  $\chi\bar{\chi}\rightarrow\mu\bar{\mu}$ and $\chi\bar{\chi}\rightarrow b\bar{b}$  for $m_\chi = $100 GeV, under the same assumptions made by the Fermi-LAT collaboration.  These regions were found  by varying the transition temperature of  each of the two non-standard cosmological models between 5 MeV and 3.5 GeV.  Each transition temperature provides a unique contour in the $A$ versus cross section space that gives the right relic DM density.  The Fermi-LAT bounds were then applied along each of these contours, dividing it into a section with acceptable annihilation rates and another with  rejected annihilation rates.  These contours were then combined to create the ruled out regions  seen in Fig. \ref{fig:6}.

In the limit of symmetric dark matter, i.e. $A=0$, Eq. \eqref{gfratio} reduces to $ \langle \sigma_{\chi{\bar{\chi}}}  v \rangle \leq 2\langle \sigma v \rangle_{\rm Fermi}$.  This corresponds to $ \langle \sigma_{\chi{\bar{\chi}}}  v \rangle \leq 1.5 \times 10^{-7}\,{\rm GeV}$ and $ \langle \sigma_{\chi{\bar{\chi}}}  v \rangle \leq 1 \times 10^{-8}\,{\rm GeV}$ for $\chi \bar{\chi}\rightarrow \mu \bar{\mu}$ and $\chi \bar{\chi}\rightarrow b \bar{b}$, respectively, for both the kination and scalar-tensor models.  Although they can not be read from the plot, the transition temperatures corresponding to the maximum annihilation cross sections mentioned must be  $T_{tr}\simeq 15\,{\rm MeV}$ ($100\, {\rm MeV}$) for $\chi \bar{\chi}\rightarrow \mu \bar{\mu}$, and $T_{tr}\simeq 500\,{\rm MeV}$ ($1.5\,{\rm GeV}$) for $\chi \bar{\chi}\rightarrow b \bar{b}$ in the kination (scalar-tensor) model so that $\chi$ and $\bar{\chi}$ make up all the DM.

The limiting cross sections for non-zero values of $A$ can be read off of Fig. \ref{fig:6}.  For example, if $A=3\times 10^{-12}$, the corresponding limits are $ \langle \sigma_{\chi{\bar{\chi}}}  v \rangle \leq 3 \times 10^{-7}\,{\rm GeV}$ and $ \langle \sigma_{\chi{\bar{\chi}}}  v \rangle \leq 2 \times 10^{-8}\,{\rm GeV}$ for $\chi \bar{\chi}\rightarrow \mu \bar{\mu}$ and $\chi \bar{\chi}\rightarrow b \bar{b}$, respectively.  The corresponding transition temperatures are $T_{tr}\simeq 9\,{\rm MeV}$ ($T_{tr}\simeq 55\,{\rm MeV}$) and $T_{tr}\simeq 280\,{\rm MeV}$ ($T_{tr}\simeq 800\,{\rm MeV}$) for the kination (scalar-tensor) model.

\section{Summary and Conclusions}

In this paper we have studied asymmetric dark matter which decouples from the thermal bath  during a  non-standard pre-BBN cosmological phase.  We have assumed that the dark matter particle asymmetry $A$ is created prior to the period that we consider and that subsequently the only reactions which change particle and anti-particle numbers are  pair annihilation into  Standard Model particles and the inverse reactions; thus $A$ is constant. We considered two non-standard pre-BBN models, kination and scalar-tensor models, in which the expansion rate of the Universe is respectively faster ($\sim T^3$) and slower ($\sim T^{1.2}$) than in the standard radiation dominated cosmology  ($\sim T^2$) until a transition temperature $T_{tr}$, which must be larger than about 4 MeV (the latest bound is 3.2 MeV \cite{Hannestad:2004px}) to preserve BBN and all the subsequent history of the Universe.

We find that for a range  of pair annihilation cross sections between about $5\times 10^{-6}$ GeV$^{-2}$ and the standard symmetric thermal dark matter cross section $1.8 \times 10^{-9}$ GeV$^{-2}$, the ratio of the relic abundance of the minority and majority dark matter components is highly dependent on the pre-BBN cosmology. For each annihilation cross section in this range, the smallness of this ratio in non-standard cosmological scenarios with particular $T_{tr}$ values can be compensated and even overcompensated by the increased annihilation cross section so that the annihilation rate at present of asymmetric dark matter, contrary to general  belief, could be even larger than that of  symmetric dark matter in the standard cosmology. In the standard pre-BBN cosmology the annihilation rate of asymmetric dark matter is always negligible, because as soon as the annihilation cross section increases with respect to the value which would yield equal minority and majority relic densities, even by a factor of a few, the minority component density becomes exponentially smaller than the majority density.

 Thus, if the annihilation cross section of the asymmetric dark matter candidate is known, the annihilation rate at present, if detectable,  could be used to test the Universe before Big Bang Nucleosynthesis, an epoch from which we do not  yet have  any data.

Finally, we would like to mention that there are bounds on the mass of bosonic asymmetric DM coming from its accumulation in neutron stars, potentially leading to the formation of black holes.\cite{bs}.  These bounds are weakened if the asymmetric DM has a non-negligible present annihilation rate, as presented in this paper.  In any event, although we have taken $g_{\chi}=1$ (which corresponds to bosonic DM) in Eqs. \eqref{nxeq} and \eqref{nxbareq}, the numerical results in this paper apply to fermionic DM up to corrections of $\mathcal{O} (1)$.

\acknowledgments

 GG and JH were supported in part by DOE grant DE-FG03-91ER40662, Task C.
 JH is also partly supported by the Consolider-Ingenio 2010 Programme under grant MultiDark CSD2009-00064.

\end{document}